\documentclass[aps,pre,showpacs,twocolumn]{revtex4-1}

\usepackage{amsmath,amssymb,amsfonts}
\usepackage{graphicx,epsf}
\usepackage{xspace}
\usepackage{textcomp}
\usepackage{color}
\usepackage[normalem]{ulem}

\usepackage[colorlinks,bookmarks=false,citecolor=blue,linkcolor=blue,urlcolor=blue,hyperfootnotes=true]{hyperref}

\newcommand{\be}{\begin{equation}}
\newcommand{\ee}{\end{equation}}

\newcommand{\p}{\partial} 
\newcommand{\vx}{\vec{x}}

\newcommand{\bq}{{\bf q}}

\newcommand{\bx}{{\bf x}}
\newcommand{\vq}{{\vec{q}}}
\newcommand{\vp}{{\vec{p}}}
\newcommand{\td}{\text{\tiny $D$}}
\newcommand{\xx}{\text{\tiny $X$}}
\newcommand{\yy}{\text{\tiny $Y$}}
\newcommand{\xn}{\text{\tiny $0$}}

\newcommand{\tih}{\tilde{h}}
\newcommand{\tj}{\tilde{j}}

\newcommand{\tp}{\hat{p}}
\newcommand{\tq}{\hat{q}}

\newcommand{\tg}{\hat{g}}
\newcommand{\tf}{\hat{f}}
\newcommand{\tI}{\hat{I}}

\newcommand{\xa}{\text{\tiny $\parallel$}}
\newcommand{\xe}{\text{\tiny $\perp$}}

\newcommand{\ew}{\text{\tiny $\mathrm{EW}$}}

\newcommand{\etad}{\eta^{\td}}
\newcommand{\etax}{\eta^{\xx}}

\begin{document}

\title{Strong coupling phases of the anisotropic Kardar-Parisi-Zhang equation}

\author{Thomas Kloss$^1$, L\'eonie Canet$^{2}$, and  Nicol\'as Wschebor$^{3}$}
\affiliation{
$^1$\mbox{IIP, Universidade Federal do Rio Grande do Norte, Av.\ Odilon Gomes de Lima  1722, 59078-400 Natal, Brazil}\\
$^2$ Laboratoire de Physique et Mod\'elisation des Milieux Condens\'es,\\
  Universit\'e Joseph Fourier  and CNRS, 25, avenue des Martyrs, BP 166, F-38042 Grenoble, France\\
$^3$Instituto de F\'isica, Facultad de Ingenier\'ia, Universidad de la Rep\'ublica, J.H.y Reissig 565, 11000 Montevideo, Uruguay}

\begin{abstract}

We study the anisotropic Kardar-Parisi-Zhang equation using nonperturbative
renormalization group methods.  In contrast to a previous analysis in the weak-coupling regime, we find the strong coupling fixed point corresponding to the isotropic rough phase to be always locally stable and unaffected by the anisotropy even at non-integer dimensions. Apart from the well-known weak-coupling and the now well-established isotropic strong coupling  behavior, we find an anisotropic strong coupling fixed point for nonlinear couplings of opposite signs at non-integer dimensions.

\end{abstract}
\date{December 22, 2014}
\maketitle

\section{Introduction}

The Kardar-Parisi-Zhang (KPZ) equation \cite{kardar86,halpin-healy95,krug97,Barabasi95}, originally intended to describe growing interfaces, has become a paradigm for kinetic roughening and dynamical scaling phenomena. Apart from its relevance in statistical physics  a connection to condensed matter systems has also been established recently \cite{Kulkarni13,Gladilin14}.
A variant of the original KPZ model, namely the anisotropic Kardar-Parisi-Zhang (AKPZ) equation was introduced some years later in the 90s by Villain and Wolf \cite{Villain91,Wolf91}, aiming at describing vicinal surfaces and the effect of anisotropy. Being for a long time more a subject of academic interest and much less studied than the standard KPZ model, the AKPZ equation has also experienced recently an unexpected relevance in condensed matter systems.
It was shown by Chen and coworkers that the order parameter for the dynamics of a driven Bose-Einstein condensate maps onto a compact form of the AKPZ equation \cite{Chen13,Altman14}. 
In contrast to the standard KPZ equation however, where by means of an exact solution in one dimension \cite{Praehofer04,Sasamoto05,Sasamoto10,Calabrese11,Amir11,Imamura12,Corwin12} and  extensive simulations in higher dimensions 
\cite{Tang92b,Ala-Nissila93,Castellano99,Marinari00,Reis04,Ghaisas06,Kelling11,Pagnani13,Halpin-Healy12,Halpin-Healy13,*Halpin-Healy13err,Alves14}
the phase diagram is relatively well understood, much less effort was spent on the AKPZ equation and  most of the studies are  limited to the weak-coupling regime \cite{Wolf91,taeuber02,Chattopadhyay07,Borodin09,Borodin14,Vivo14}.
An understanding of the AKPZ equation which includes the strong-coupling behavior was missing until now.

The AKPZ equation assumes the $d$-dimensional space to be partitioned into two orthogonal and isotropic subspaces with dimensions $d_\xe$ and $d_\xa$.
In the original work by Wolf \cite{Wolf91} where the dimensions of the two sectors were chosen equally ($d_\xa = d_\xe = 1$) so that the total dimension is $d = 2$, the anisotropy is found to be only of minor importance and basically leading to the same  universality classes  as  the isotropic KPZ equation.
That is, if both couplings were chosen with an equal sign, the isotropic KPZ behavior was recovered at long distances whereas in the case of an opposite sign or with one vanishing coupling the weak-coupling Edwards-Wilkinson (EW) behavior with logarithmic roughness was obtained. The results of Wolf were confirmed by numerical simulations \cite{Halpin-Healy92,Jeong93,Kim98a} and the logarithmic correlations by exact arguments \cite{Praehofer97}.
In contrast, studies on variants of the KPZ equation \cite{Jeong95,Mukherji97,Jung00,Tang02,Tang02a}
as well as on related models \cite{hwa92,Tang95} indicated that anisotropy may become important in  some cases.
A subsequent study of the AKPZ equation performed by T\"auber and Frey (TF) \cite{taeuber02} by means of dynamic renormalization group later revealed that the result of Wolf is only a special case and the situation in generic dimension is  more complex.
Depending on the total dimension $d$ and on the splitting in the two sector dimensions $d_\xa$ and $d_\xe$, TF argued that the anisotropy can lead not only to new universality classes but may also cause the isotropic solution to become unstable.

In the present study, we reexamine the AKPZ equation using  nonperturbative renormalization group (NPRG) methods. The NPRG approach to the KPZ equation was developed in \cite{Canet10,Canet11a,*Canet12Err,Kloss12,Kloss14}.
It successfully yields the KPZ phase diagram with the correct strong coupling behavior and the corresponding exponents are in close  agreement
with simulation results in $d=2$ and $d=3$. In $d = 1$  the exact exponents are recovered and the scaling function of the height-height correlation compares very accurately with the exact scaling function from Ref.\ \cite{Praehofer04}.
Throughout this work we use two different approximations: the simpler local-potential prime approximation (LPA') \cite{Delamotte12,*delamotte07} to obtain the phase diagram and the more accurate Next-to-Leading Order (NLO) approximation of Ref.\ \cite{Kloss12}  in order to check our findings.
The LPA' approximation does not lead to accurate numbers for the exponents but in general it reproduces qualitatively the phase diagram with the correct strong coupling behavior. A similar strategy with these two approximations has recently been successfully followed to study the KPZ equation with Gaussian long-range correlated noise \cite{Kloss14}.

As shown by TF, the stability of the weak-coupling isotropic EW fixed point as well as of the weak-coupling uniaxial  (EWU) fixed point are changed when a third weak-coupling anisotropic (EWA) fixed point is crossing them. This happens in certain dimension combinations.
Their analysis also applies the transition fixed points in the leading $\epsilon = d - 2$ expansion.
They conjectured that this perturbative analysis of the stability of the KPZ equation against anisotropic perturbations could be extended to the transitions points for more general dimensions and probably to the strong coupling regime.
TF further speculated that a finite upper critical dimension may result from this instability.
Our analysis of the NPRG flow equations however shows that no prediction for the strong-coupling behavior can be drawn from the stability changes in the weak-coupling sector.
In the weak coupling regime, our results are in close agreement with the ones by TF and we recover the same fixed points with the predicted stability changes. 
In addition to the weak coupling regime, the NPRG approach also describes  the strong coupling behavior with the isotropic rough phase, characterized by a fully attractive strong coupling (SC) fixed point. If both initial nonlinear couplings are chosen with an equal sign (and, in the case $d > 2$ sufficiently large) the isotropic strong-coupling fixed point is always reached. That is, we find that the stability change of the isotropic EW fixed point, triggered by the crossing EWA fixed point does not affect the isotropic strong-coupling fixed point. 
In our scenario the isotropic rough phase is always locally stable for an arbitrary splitting in the two sector dimensions. If the initial nonlinear couplings are chosen with  opposite signs or if one of them is zero (the uniaxial case) additional strong-coupling fixed points are found in some cases. We observe however, that these fixed points do not enter in the quadrant where both couplings have the same sign and therefore never cross the isotropic strong-coupling fixed point to change its stability.

The paper is organized as follows. In Sec.\ \ref{NPRG}, we shortly present the NPRG formalism for the AKPZ equation, our ansatz with the approximations used and the corresponding 
flow equations. In Sec.\ \ref{sec:results} the fixed point solutions of the flow equations are presented and the full phase diagram of the system is determined. The results are summarized in Sec.\ \ref{sec:conclusion}. Technical details and the connection to the results by TF can be  found in the appendices.

\section{Method}
\label{NPRG}

\subsection{AKPZ field theory}

The AKPZ equation \cite{Wolf91} reads 
\begin{align}
& \frac{\p h(t,\vx)}{\p t} = \nu_{\parallel} \,\nabla^2_{\parallel} h(t,\vx) \, + \nu_{\perp} \,\nabla^2_{\perp} h(t,\vx) \, \nonumber \\
& \,\, +
\,\frac{\lambda_{\parallel}}{2}\,\big(\nabla_{\parallel}  h(t,\vx)\big)^2 \,+
\,\frac{\lambda_{\perp}}{2}\,\big(\nabla_{\perp}  h(t,\vx)\big)^2 \,+\,\eta(t,\vx) ,
\label{eq:akpz}
\end{align}
where $h(t,\vx) \equiv h(t,\vx_\xa,\vx_\xe)$ is a single valued height profile which depends on the spatial substrate coordinates $\vx= (\vx_{\parallel} ,\vx_{\perp})^T$ and on time $t$. Parallel and perpendicular components of the substrate have
 dimensions $d_{\xa}$ and $d_{\xe}$ in each sector so that the total substrate dimension is $d = d_{\parallel}  + d_{\perp}$.
The AKPZ equation has two surface-tension terms and two non-linear couplings proportional to $\nu_{i}$ and $\lambda_{i}$ with $i = \{ \parallel, \perp \}$. The last term 
 $\eta(t,\vx)$ represents the noise. As in Refs.\ \cite{Wolf91,taeuber02} we choose  Gaussian white noise with zero mean $\langle \eta(t,\vx)\rangle = 0$ and the isotropic correlator 
\begin{equation}
  \big\langle \eta(t,\vx)\eta(t',\vx')\big\rangle = 2 D\, \delta^{(d_{\xa})}(\vx_{\xa}-\vx'_{\xa})\, \delta^{(d_\xe)}(\vx_{\xe}-\vx_{\xe}')\, \delta(t-t') ,
\end{equation}
where $D$ is the noise amplitude.
A field theory can be derived in the usual way with the Janssen-de Dominicis procedure \cite{janssen76, *dominicis76}, introducing an additional Martin-Siggia-Rose response field $\tih$ \cite{Martin73}. Eq.\ (\ref{eq:akpz})  thereby yields the generating functional  
\begin{subequations}
\begin{align}
\!\!\!\!{\cal Z}[j,\tj] \! &= \!\!\int \!{\cal D}[h,i \tih]\, 
\exp \! \left(-{\cal S}[h,\tih] +  \int_{\bx} \left\{ j h+\tj\tih\right\} \right) , \label{Z}\\
\!\!\!\!{\cal S}[h,\tih]  \! &= \!\! \int_{\bx}  \!\bigl\{ \tih(\bx)\bigl(\p_t h(\bx) -\nu_{\parallel}  \,\nabla^2_{\parallel} h(\bx) -\nu_{\perp}  \,\nabla^2_{\perp} h(\bx) \nonumber \\
& \!\!\!\! \!\!\!\! \!\!\! - 
\frac{\lambda_{\parallel}}{2}\,\left({\nabla_{\parallel}} h(\bx)\right)^2 - 
\frac{\lambda_{\perp}}{2}\,\left({\nabla_{\perp}} h(\bx)\right)^2\bigr) - D (\tih(\bx))^2 \bigr\}  ,
\label{eq:action}
\end{align} 
\end{subequations}
where $j$ and $\tj$ are sources and we have introduced the notation $\bx \equiv (t,\vec{x}_{\xa},\vec{x}_{\xe})$ for convenience. For later reference, we also write frequency and momentum labels as $\bq \equiv (\omega,\vec{q}_{\xa},\vec{q}_{\xe})$ and use the Fourier transform convention of Ref.\ \cite{Kloss12} throughout this work. Being only of technical relevance and without loss of generality, we work in the following with the rescaled AKPZ action,
such that $\nu_{\xa} = \nu_{\xe}$, see appendix B.

The AKPZ action Eq.\ (\ref{eq:action}) is invariant under both infinitesimal field transformations
\begin{subequations} 
\begin{eqnarray}
\text{(i)} && \left\{
\begin{array}{l}
h'(t,\vx)=\vx \cdot \p_t \vec v(t) + h(t,\vx+ \langle \vec{\lambda} , \vec v(t) \rangle )\label{galg}\\
\tilde h'(t,\vx)=\tilde h(t,\vx+ \langle \vec{\lambda} ,\vec v(t) \rangle)
\end{array}
\right.\\
\text{(ii)} && \;\;\;\; h'(t,\vx)=h(t,\vx)+c(t), \label{timeg}
\end{eqnarray}
\label{eq:sym}
 \end{subequations}
with $\vec v \equiv (\vec v_\xa , \vec v_\xe)^T$ and $\langle \vec{\lambda} , \vec{v} \rangle \equiv  (\lambda_\xa \vec v_\xa , \lambda_\xe \vec v_\xe)^T$ apart from some terms which variations are linear in the fields.
The parameters $\vec v$ and $c$ of these transformations are infinitesimal vector and scalar respectively,  depending on time.
Relation (i) corresponds to the Galilean symmetry in a time-gauged form \cite{Canet11a,Lebedev94} and (ii) is the shift symmetry also gauged in time. As in reference \cite{Canet11a}, general Ward identities can be deduced from them. These symmetries  constitute the central ingredients to device our NPRG ansatz in Sec.\ \ref{sec:ansatz}.
Note that the Cole-Hopf version of the AKPZ action with the related Z$_2$ symmetry holds only in the specific case where $\nu_\xa \lambda_\xe / (\nu_\xe \lambda_\xa) = 1$ as pointed out in Ref.\ \cite{Wolf91}.
Contrarily to the isotropic situation, there is no time reversal symmetry \cite{Frey96,Canet10} in the anisotropic case.
In the symmetric case where $\nu_\xa = \nu_\xe$ and $\lambda_\xa = \lambda_\xe$ an additional rotational symmetry is present and the AKPZ action identifies with the isotropic one. 
In the uniaxial case with $\lambda_\xe = 0$ an additional symmetry  exists and this implies that the hypersurface $\gamma = \nu_\xa / \nu_\xe = 0$ is closed along the flow. To the best of our knowledge this symmetry has not yet been discussed elsewhere in the literature. We refer to appendix E for further details.

\subsection{Nonperturbative renormalization group}

The NPRG can be seen as a generalization of Wilson's idea to construct an effective field theory for the long-range physics by continuously integrating out fast degrees of freedom. We do not go into details here and only highlight  the main steps.  We refer the reader  to Refs.\ \cite{Berges02,Delamotte12,*delamotte07,Kopietz10} for a general introduction to the NPRG. The implementation of the NPRG method to describe out-of-equilibrium situations can be found e.g.\ in Refs.\ \cite{canet04a,Canet11b,Kloss11,Berges12} and the first application to the KPZ problem is achieved in Ref.\  \cite{Canet10}.
The starting point in a NPRG description is to add a regulator term
\begin{equation}
\Delta {\cal S}_\kappa \!=\!\frac{1}{2}\! \int_{\bf q}\!  h_i(-{\bf q})\, 
[R_\kappa({\bf q})]_{ij}\, h_j({\bf q}) ,\;\;\label{deltask}
\end{equation}
to the action Eq.\ (\ref{eq:action}), where $i,j\in\{1,2\}$ label the field and response field $h_1=h,h_2=\tih$. $R_\kappa$ is the regulator matrix and $\kappa$ is a running cutoff. The generating functional therefore depends on the external scale $\kappa$
\begin{equation}
{\cal Z}_\kappa[j,\tj] \!\! = \!\!\!\int {\cal D}[h,i \tih]\, 
\exp\left(-{\cal S}-\Delta{\cal S}_\kappa+  \int_{\bf x} \left\{ j h+\tj\tih \right\}\right).  \label{Zk}
\end{equation} 
The form of the regulator is quite general but its asymptotic behavior must fulfill some requirements. For $\kappa \gg q$ the regulator behaves like a mass with $R_\kappa \sim \kappa^2$ so that all fluctuations are suppressed in the limit $\kappa \rightarrow \infty$. Lowering the cutoff down to $\kappa \ll q$ the cutoff function vanishes, so that in the limit $\kappa \rightarrow 0$ the original theory without regulator is recovered.
We choose the regulator function as
\begin{equation}
R_\kappa({\bf q}) \!=\! r\left(\frac{q^2}{\kappa^2}\right)
\left(\!\! \begin{array}{cc}
0&  {\nu^\xa_\kappa q_{\xa}^2  + \nu^\xe_\kappa q_{\xe}^2} \\
 {\nu^\xa_\kappa q_{\xa}^2  + \nu^\xe_\kappa q_{\xe}^2} & -2 D_\kappa 
\end{array}\!\!\right) ,
\label{eq:Rk}
\end{equation}
with $q_i = |\vec{q}_i|$ where  $i = \{ \parallel, \perp \}$ such that it preserves all the symmetries of the AKPZ action. The scale dependent parameters $\nu^\xa_\kappa, \nu^\xe_\kappa$ and $D_\kappa$, are defined later in Eqs.\ (\ref{eq:etaflow},\ref{eq:coeff}).
We use an exponential cutoff function of the form
\begin{equation}
r(x)=\alpha/(\exp(x) -1). 
\label{eq:expReg}
\end{equation}
where $\alpha$ is a free parameter,  which can be varied to assess the quality of an approximation. Unless otherwise indicated, we choose $\alpha = 4$ throughout this work since the variations of the critical exponents are minimal around this value for $d = 2, 3$ \cite{Kloss14}.
Defining the functional  ${\cal W}_\kappa = \log {\cal Z}_\kappa$,
field expectation values are obtained as functional derivatives with respect to the sources $j$ and $\tilde{j}$ 
\begin{equation}
  \varphi({\bf x}) = \langle h({\bf x}) \rangle = \frac{\delta {\cal W}_{\kappa}}{\delta j({\bf x})}  \, \, , \, \,
  \tilde \varphi({\bf x}) = \langle \tilde h({\bf x}) \rangle = \frac{\delta {\cal W}_{\kappa}}{\delta \tilde j({\bf x})}  .
\end{equation} 
The effective action $\Gamma_\kappa[\varphi,\tilde\varphi]$ is defined as 
\begin{equation}
\Gamma_\kappa[\varphi,\tilde\varphi] +{\cal W}_\kappa[j,\tj] = 
\int \! j_i \varphi_i -\frac{1}{2} \int \varphi_i \, [R_\kappa  ]_{ij}\, \varphi_{j} ,
\label{legendre}
\end{equation}
which is up to a term proportional to $R_\kappa$ the Legendre transform of  ${\cal W}_\kappa$.
The exact flow for $\Gamma_{\kappa}[\varphi,\tilde\varphi]$ is given  by Wetterich's 
equation, which  reads in Fourier space  \cite{Wetterich93,Berges02,Delamotte12,*delamotte07,Kopietz10}
\begin{equation}
\partial_\kappa \Gamma_\kappa = \frac{1}{2}\, {\rm Tr}\! \int_{\bf q}\! \partial_\kappa R_\kappa \cdot G_\kappa ,
\label{eq:dkgam}
\end{equation}
and where
\begin{equation}
 G_\kappa=\left[\Gamma_\kappa^{(2)}+R_\kappa\right]^{-1} ,
 \label{eq:propag}
\end{equation}
is the full propagator in presence of external fields.

\subsection{Ansatz}
\label{sec:ansatz}

 Our aim is to build  an approximation scheme for the AKPZ equation, which automatically preserves along the flow the symmetries of the problem, summarized in Eq.\ (\ref{eq:sym}).
The strategy which is inspired from Refs.\ \cite{blaizot06,Benitez12} and is similar to the approach used in Ref.\ \cite{Canet11a,*Canet12Err} for the isotropic KPZ equation, is to construct an ansatz in terms of symmetry invariant building blocks.
In the AKPZ equation these building blocks are the covariant time derivative of the field
\begin{equation}
D_t \varphi  = \partial_t \varphi - \frac{\lambda_\xa}{2} (\nabla_{\xa} \varphi)^2 - \frac{\lambda_\xe}{2} (\nabla_{\xe} \varphi)^2 ,
\end{equation}
and the  Galilean invariants 
$\tilde \varphi$, $\nabla^2_\xa \varphi$, $\nabla^2_\xe \varphi$ and gradients and appropriate covariant time derivatives of them. In order to keep the full momentum dependence of the 2-point functions, we further introduce four running functions $f^\xx_\kappa(\vp)$ with $X \in \{ \lambda, D, \nu_\xa, \nu_\xe \}$. In principle, these running functions also depend on the frequency like in Ref.\ \cite{Canet11a,*Canet12Err}. However, to preserve the anisotropic momentum dependence while keeping the equations numerically tractable, we use the next-to-leading order (NLO) approximation of Ref.\ \cite{Kloss12}.  It consists in neglecting the frequency dependence of the four flowing functions $f_\kappa^\xx(\omega,\vp)\to f_\kappa^\xx(\vp)$  within the loop integrals which reduces drastically the numerical complexity. In the isotropic case, this approximation is reliable typically up to $d\lesssim 3.5$,  as can be inferred from the dependence of the exponents in the cutoff parameter $\alpha$, see \cite{Kloss12}. Moreover, we only focus in this work on the zero frequency sector and thus the  frequency dependence can be completely dropped from the running functions such that the NLO ansatz for the AKPZ equation reads 
\begin{align}
\Gamma_\kappa[\varphi,\tilde \varphi]\!  =&   \! \!  \int_{\bf x} \!
\Bigl\{ \tilde \varphi f_\kappa^\lambda(\nabla) D_t \varphi  - \ (\nabla^2_{\xa} \varphi) f_\kappa^{\nu \xa}(\nabla) \tilde \varphi \nonumber \\
  &\hspace*{2ex}  - (\nabla^2_{\xe} \varphi) f_\kappa^{\nu \xe}(\nabla) \tilde \varphi   - 
\tilde \varphi f_\kappa^\td(\nabla) \tilde \varphi  \Bigr\},
\label{anznlo}
\end{align}
with $f_\kappa^\xx(\nabla) \equiv  f_\kappa^\xx(-\nabla_{\xa}^2,-\nabla_{\xe}^2)$.
Taking functional derivatives of the ansatz w.r.t.\ $\varphi$ and $\tilde{\varphi}$ and evaluating them at $\varphi = \tilde{\varphi} = 0$ we obtain for the 2-point functions
\begin{subequations}
\begin{align}
\Gamma_\kappa^{(2,0)}(\omega,\vec{p}_{\xa},\vec{p}_{\xe}) &= 0 , \\
\Gamma_\kappa^{(1,1)}(\omega,\vec{p}_{\xa},\vec{p}_{\xe}) &= i \omega \,f_\kappa^\lambda\left( p_{\xa},p_{\xe} \right) +  \vp\,^2_\xa  f_\kappa^{\nu \xa}(p_{\xa},p_{\xe}) \nonumber \\ 
&  +  \vp\,^2_\xe  f_\kappa^{\nu \xe}(p_{\xa},p_{\xe}) , \\
\Gamma_\kappa^{(0,2)}(\omega,\vec{p}_{\xa},\vec{p}_{\xe}) &= -2  f_\kappa^\td(p_{\xa},p_{\xe}).  
\end{align}
\label{eq:anzgam2}
\end{subequations}
Further, we make the reasonable approximation
\begin{equation}
f_\kappa^{\nu \xe}(p_{\xa},p_{\xe}) = \gamma_\kappa^{-1} f_\kappa^{\nu \xa}(p_{\xa},p_{\xa}) ,
\end{equation}
where $\gamma_\kappa= \nu_\kappa^{\xa} / \nu_\kappa^{\xe}$ and use the notation $f_\kappa^{\nu \xa} \equiv f_\kappa^{\nu}$ for convenience. We are left with three flowing functions $f_\kappa^\xx$ where $X \in \{ \lambda, D, \nu \}$.
These functions are extracted from the 2-point functions as
\begin{subequations}
\begin{eqnarray}
f_\kappa^\nu(p_{\xa},p_{\xe})  &=&  \left. \frac{1}{\vp_{\xa}^2 +  \gamma_\kappa^{-1} \vp_{\xe}^2}   \Re \Gamma_\kappa^{(1,1)}(\omega,\vec{p}_{\xa},\vec{p}_{\xe})  \right|_{\omega = 0} , \\
f_\kappa^\lambda\left(p_{\xa},p_{\xe} \right) &=& \left. \frac{1}{\omega } \Im  \Gamma_\kappa^{(1,1)}(\omega,\vec{p}_{\xa},\vec{p}_{\xe})  \right|_{\omega = 0}, \\
f_\kappa^\td(p_{\xa},p_{\xe})  &=& - \left. \frac{1}{2}  \Gamma_\kappa^{(0,2)}(\omega,\vec{p}_{\xa},\vec{p}_{\xe}) \right|_{\omega = 0} .
\end{eqnarray}
\end{subequations}
From Eqs.\ (\ref{eq:Rk},\ref{eq:propag},\ref{eq:anzgam2}) the propagator follows as
\begin{equation}
G_\kappa(\omega,\vq_\xa,\vq_\xe) =\frac{1}{P_\kappa(\omega,q)}\left(\!\! \begin{array}{cc}
2 k_\kappa(q) &  Y_\kappa(\omega,q_\xa, q_\xe)\\
 Y^*_\kappa(\omega,q_\xa, q_\xe) & 0 
\end{array}\!\!\right) , \label{propag}
\end{equation}
where
\begin{subequations}
\begin{align}
k_\kappa( q_{\xa},q_{\xe}) &= f^\td_\kappa(q_{\xa},q_{\xe})+ D_\kappa \,r( q^2/\kappa^2) , \\
l_\kappa( q_\xa, q_\xe) &=  q_{\parallel}^2  ( f^\nu_\kappa\left( q_{\xa},q_{\xe}\right)+  \nu^\xa_\kappa\, r( q^2/\kappa^2) ) \nonumber \\
&+  q_{\xe}^2  ( f^\nu_\kappa\left( q_{\xa},q_{\xe}\right)+  \nu^\xe_\kappa\, r( q^2/\kappa^2) ) , \\
Y_\kappa(\omega, q_\xa, q_\xe)&= i \omega \,f^\lambda_\kappa(q_{\xa},q_{\xe})+ l_\kappa( q_\xa, q_\xe)  , \\
P_\kappa(\omega, q_\xa, q_\xe)& = (\omega \,f^\lambda_\kappa \! \left(q_{\xa},q_{\xe}\right))^2 +  (l_\kappa\! \left(q_\xa, q_\xe\right))^2 .
\end{align}
\end{subequations}
The scale derivative of the regulator matrix Eq.\ (\ref{eq:Rk})  is
\begin{align}
\label{eq:dRk}
&\partial_\kappa R_\kappa(\vq) = \nonumber \\
&\left(\!\! \begin{array}{cc}
0                                                 &      { q^2_{\parallel} \partial_\kappa S^{\xa}_\kappa (q)  +  q^2_{\perp} \partial_\kappa S^{\xe}_\kappa (q)} \\
{ q^2_{\parallel} \partial_\kappa S^{\xa}_\kappa (q)  +  q^2_{\perp} \partial_\kappa S^{\xe}_\kappa (q) } &  -2  { \partial_\kappa S^\td_\kappa (q)}
\end{array}\!\!\right) ,
\end{align}
where 
  \begin{equation}
S_\kappa^\xx (q^2)  =  X_\kappa r(y)  \, \, ,\, \, y = q^2/\kappa^2  ,
\end{equation}
and
\begin{equation}
\kappa \partial_\kappa S^\xx_\kappa (y)  = - X_\kappa \, (\etax_\kappa r(y) + 2 y \,\partial_{y}  r(y)).
\label{eq:kdkS} 
\end{equation}
and we use the notation
\begin{equation}
X \in \{ D,\parallel , \perp \} ,  \quad X_\kappa \in \{ D_\kappa,\nu^\xa_\kappa , \nu^\xe_\kappa \} , \quad \etax_\kappa \in \{ \etad_\kappa,\eta^\xa_\kappa , \eta^\xe_\kappa \} .
\end{equation}
The running coefficients $X_\kappa$ and anomalous dimensions $\etax_\kappa$  will be defined later in Eq.\ (\ref{eq:etaflow}).
Within the NLO approximation, the remaining $n$-point functions all vanish, except
\begin{align}
&\Gamma_\kappa^{(2,1)}(\omega_1,\omega_2;\vp_{\parallel 1},\vp_{\parallel 2} ,\vp_{\perp 1},\vp_{\perp 2}) = \nonumber \\
&  (\lambda_{\parallel} \, \vp_{\parallel 1}\cdot \vp_{\parallel 2} + \lambda_{\perp} \, \vp_{\perp 1}\cdot \vp_{\perp 2} )\, f_\kappa^\lambda\left(|\vp_{\parallel 1}+\vp_{\parallel 2} |, |\vp_{\perp 1}+\vp_{\perp 2} |\right). 
\label{gam3aniso}
\end{align}

\subsection{Dimensionless flowing functions}
To analyze the fixed-point properties of the flow equations, dimensionless and renormalized quantities (indicated by a hat) are introduced. Momentum and frequency are measured in units of the running cutoff $\kappa_\xa \equiv \kappa$ 
 \begin{equation} 
 \hat{p}_\xa = p_\xa / \kappa , \quad \hat{p}_\xe =  p_\xe / ( \gamma^{1/2}_\kappa \kappa )    , \quad  \hat{\omega} = \omega / (\nu^\xa_\kappa \kappa^2) ,
\label{eq:dimvar}
 \end{equation} 
 and we define three running coefficients $\{ D_\kappa,\nu_\kappa^{\xa},\nu_\kappa^{\xe} \}$ and the related anomalous dimensions $\etax_\kappa$  via
 \begin{equation} 
 \etad_\kappa = -\kappa \partial_\kappa \ln D_\kappa  , \quad
 \eta^\xa_\kappa = -\kappa\partial_\kappa\ln \nu^\xa_\kappa , \quad \eta^\xe_\kappa = -\kappa\partial_\kappa\ln \nu^\xe_\kappa .
  \label{eq:etaflow}  
  \end{equation}
At the initial cutoff scale $\kappa \equiv \Lambda$ the running coefficients are equal to unity: $D_\Lambda = \nu^{\xa/\xe}_\Lambda = 1$. Due to the Galilean symmetry Eq.\ (\ref{galg}) both coefficients $\lambda_{\xa/\xe}$ are not renormalized and thus no additional running couplings are needed for them.
We define the anisotropy $\xi$, dynamical $z$ and roughness  $\chi$ exponents as
 \begin{equation} 
x_\xe \sim x_\xa^\xi , \quad t \sim x_\xa^z , \quad h \sim x_\xa^\chi .
\label{eq:anisoexp}
\end{equation} 
With the field dimensions $[h(\bx)] = (\kappa^{d-2} D_\kappa \gamma^{d_\perp / 2}_\kappa / \nu^\xa_\kappa)^{1/2}$ and $[\tilde{h}(\bx)] = (\kappa^{d+2} \nu^\xa_\kappa \gamma^{d_\perp / 2}_\kappa / D_\kappa )^{1/2}$ and the rescaling Eq.\ (\ref{eq:dimvar}),  the physical critical exponents can be deduced from the anomalous dimensions at a fixed point (indexed by a star) via
\begin{subequations}
 \begin{align} 
\xi &= 1 - (\eta_*^\xa - \eta_*^{\xe} ) / 2  ,  \\ 
z&=2-\eta^\xa_*   ,   \\  
 \chi &= (2-d+\etad_*-\eta^\xa_* + d_\perp / 2 (\eta_*^\xa   - \eta_*^\xe ))/2. 
\label{eq:expo}
 \end{align} 
\end{subequations} 
Expressed in terms of dimensionless quantities, the AKPZ flow equations depend on the two dimensionless couplings 
\begin{subequations}
\begin{align}
\hat{g}^\xa_\kappa  & =  g^\xa_b \kappa^{d-2} \left( \frac{ D_\kappa \gamma^{d_\perp / 2}_\kappa}{\nu_{\xa \kappa}^{3}}\right) ,  \\
\hat{g}^\xe_\kappa  & = \hat{g}^\xa_\kappa \gamma^2_\kappa \frac{g^\xe_b}{g^\xa_b},
\end{align}
\end{subequations} 
where the bare couplings are defined in Eq.\ (\ref{eq:effCoupl}).
We will not study the flow in terms of the nonlinear couplings $\hat{g}^\xa_\kappa$ and $\hat{g}^\xe_\kappa$ directly, but rather in terms of $\hat{g}^\xa_\kappa  \equiv \hat{g}_\kappa $ and the anisotropy ratio $\gamma_\kappa$.
The flows of the coupling and of the anisotropy ratio are
\begin{subequations}
\begin{align}
 \partial_s \tg_\kappa &= \tg_\kappa (d - 2-\etad_\kappa + 3\eta^{\xa}_\kappa - (d_\xe / 2) (\eta^{\xa}_\kappa - \eta^{\xe}_\kappa )) ,
 \label{eq:gflow} \\
 \partial_s \gamma_\kappa &= - \gamma_\kappa (\eta^{\xa}_\kappa - \eta^{\xe}_\kappa)  ,
 \label{eq:uflow} 
 \end{align}
\label{eq:betaflow}
\end{subequations} 
where $s = \ln(\Lambda / \kappa)$ is called the RG ``time'' and  $\partial_s \equiv \kappa \partial_\kappa$.

Further, dimensionless running functions with two momentum arguments are defined by
 \begin{equation}  
 \tf_\kappa^\xx(\tp_\xa, \tp_\xe)  = f_\kappa^\xx(p_\xa,p_\xe)/X_\kappa ,
  \label{eq:dimlessFunc}  
 \end{equation}  
for $X \in \{D,\nu,\lambda\}$ and  $X_\kappa \in  \{D_\kappa,\nu^{\xa}_\kappa,1\}$.  Their flows are given by
\begin{align} 
& \partial_s \tf_\kappa^\xx(\tp_\xa, \tp_\xe) 
  = \etax_\kappa \tf_\kappa^\xx(\tp_\xa, \tp_\xe)  +  \tp_\xa \;\p_{\tp \xa} \tf_\kappa^\xx(\tp_\xa, \tp_\xe)  \nonumber \\
  &+ (1 - \eta^\xa_\kappa / 2 + \eta^\xe_\kappa / 2 ) \tp_\xe \p_{\tp \xe} \tf_\kappa^\xx(\tp_\xa, \tp_\xe) +\tI_\kappa^\xx(\tp_\xa, \tp_\xe) ,
  \label{eq:dimlessFlowf}
  \end{align}
with $\etax_\kappa \in  \{\eta^\td_\kappa,\eta^{\xa}_\kappa,0\}$,
and  $\tI_\kappa^\xx(\tp_\xa, \tp_\xe)$ are the loop integrals 
 \begin{equation}  
 \tI_\kappa^\xx(\tp_\xa, \tp_\xe) =  (\kappa \partial_\kappa  f_\kappa^\xx(p_\xa,p_\xe))/X_\kappa .
 \label{eq:loopI}
 \end{equation}  
At the origin, the scale dependent functions are normalized such that $\tf_\kappa^\xx(0,0) = 1$.

Finally, we have to calculate the flow of the anomalous dimensions from the ansatz for $\Gamma_\kappa$. 
We deduce from the 2-point functions in Eqs.\ (\ref{eq:anzgam2}) that the running coefficients can be expressed as
\begin{subequations}
  \begin{align}
\nu^\xa_\kappa &= \lim_{\omega \rightarrow 0, \vec{p}_\xa \rightarrow 0} \left( \lim_{\vec{p}_\xe \rightarrow 0} \Gamma^{(1,1)}(\omega,\vec{p}_\xa,\vec{p}_\xe) / \vec{p}_\xa^2 \right) , \\
\nu^\xe_\kappa &= \lim_{\omega \rightarrow 0, \vec{p}_\xe \rightarrow 0} \left( \lim_{\vec{p}_\xa \rightarrow 0} \Gamma^{(1,1)}(\omega,\vec{p}_\xa,\vec{p}_\xe) / \vec{p}_\xe^2 \right) , \\
D_\kappa &= - \lim_{\omega \rightarrow 0, \vec{p}_\xe \rightarrow 0, \vec{p}_\xa \rightarrow 0} \Gamma^{(0,2)}(\omega,\vec{p}_\xa,\vec{p}_\xe) / 2 .
  \end{align}
  \label{eq:coeff}
\end{subequations}
Note the ordering of the limits for the two momenta to obtain the parallel and perpendicular components.
From Eq.\ (\ref{eq:loopI}), we further define loop integrals with zero external arguments 
\begin{subequations}
  \begin{align}
\tI_\kappa^\xa &= \lim_{\omega \rightarrow 0, \vec{p}_\xa \rightarrow 0} \left( \lim_{\vec{p}_\xe \rightarrow 0} \kappa \partial_\kappa \Gamma^{(1,1)}(\omega,\vec{p}_\xa,\vec{p}_\xe) / ( \nu^\xa_\kappa \vec{p}_\xa^2) \right) , \\
\tI_\kappa^\xe &= \lim_{\omega \rightarrow 0, \vec{p}_\xe \rightarrow 0} \left( \lim_{\vec{p}_\xa \rightarrow 0} \kappa \partial_\kappa  \Gamma^{(1,1)}(\omega,\vec{p}_\xa,\vec{p}_\xe) / (\nu^\xe_\kappa \vec{p}_\xe^2) \right) , \\
\tI_\kappa^\td &= - \lim_{\omega \rightarrow 0, \vec{p}_\xe \rightarrow 0, \vec{p}_\xa \rightarrow 0} \kappa \partial_\kappa  \Gamma^{(0,2)}(\omega,\vec{p}_\xa,\vec{p}_\xe) /( 2 D_\kappa) .
  \end{align}
  \label{eq:loopints}
\end{subequations}
Therefore, from the definition of the anomalous dimensions Eq.\ (\ref{eq:etaflow}) and Eqs.\ (\ref{eq:coeff}, \ref{eq:loopints})
\begin{equation}  
  0 = \eta_\kappa^\xx  +\tI_\kappa^\xx
\label{etapert}
\end{equation}  
follows, where
$X \in  \{D , \parallel , \perp \}$.
Due to the regulator each integral $\tI_\kappa^\xx$ itself depends linearly on   $\eta_\kappa^\xx$ and can be written in the form
  \begin{equation}
     \tI_\kappa^{\xx} = \tI^{\xx \td}_\kappa \etad_\kappa + \tI^{\xx\xa}_\kappa \eta^{\xa}_\kappa + \tI^{\xx \xe}_\kappa \eta^{\xe}_\kappa  + \tI^{\xx \xn}_\kappa .
     \label{eq:etaI}
  \end{equation}
Eqs.\ (\ref{etapert}) and (\ref{eq:etaI}) thus form a linear set of equations that can be solved for the exponents. Explicit expressions for the integrals $\tI^{\xx \yy}_\kappa$ with $X, Y \in  \{D , \parallel , \perp \}$ are given in Eqs.\ (\ref{eq:etaint},\ref{eq:etadef}) of the appendix.

The NLO approximation \cite{Kloss12} noticeably reduces the complexity of the flow equations
but the loop-integrals in Eqs.\ (\ref{eq:loopI}) are still four-dimensional integrals and numerically cumbersome. However, for a qualitative picture of the phase diagram it is sufficient to consider only the flow of the scale dependent couplings and to set all flowing functions to one:
\begin{equation}  
 \tf_\kappa^\xx(\tp_\xa, \tp_\xe)  \rightarrow \tf_\kappa^\xx(0,0) \equiv 1 .
  \label{eq:simple}  
\end{equation}  
This approximation is usually referred to as Local 
Potential Approximation prime  (LPA') \cite{Delamotte12}.
For the isotropic KPZ equation the roughness exponent $\chi$  is overestimated at the LPA' level but the qualitative behavior and the correct strong-coupling physics is already obtained. In addition, the LPA' allows to study analytically the weak-coupling limit of the flow and we recover the results of TF in that limit, see appendix A.
Only in order to test the robustness of the various features of the phase diagram obtained in the LPA' approximation, we will resort to the NLO approximation. A similar strategy, with both LPA' and NLO, was adopted to study the long-range KPZ equation in a previous study \cite{Kloss14}. Hence, in the following,  the LPA' results are presented, except when they disagree with the NLO ones, which will be indicated.

\subsection{Numerical implementation}

In the LPA' approximation the two  flow equations  (\ref{eq:betaflow}) for the couplings, and in addition the three flow equations (\ref{eq:dimlessFlowf}) for the running functions at NLO,  are solved numerically by explicit Euler time stepping. 
Stable fixed point solutions are typically found for RG ``times''  $s \leq -10$.
The flowing functions $\hat{f}^\xx_\kappa(\hat{p}_\xa,\hat{p}_\xe)$ are discretized on rectangular and equidistant $\hat{p}_\xa \times \hat{p}_\xe$ grids. Between the gridpoints, they are interpolated by bi-cubic splines. Numerical integrations over the two radial momentum components and over the two angles are performed by Gauss-Legendre quadrature. Note, that eventhough the momentum grid in the two dimensionless momentum directions $\hat{p}_\xa$ and $\hat{p}_\xe$ is rectangular, the dimensionful grid is properly rescaled according to Eq.\ (\ref{eq:dimvar}) to account for the anisotropy.
In the LPA' approximation, the two-dimensional radial integrals for $\tI_\kappa^\xx$ Eqs.\ (\ref{eq:etaint}) are reduced to one-dimensional integrals according to Eq.\ (\ref{eq:radialint}). We initialize all flowing functions at $\hat{f}^\xx_\Lambda(\hat{p}_\xa,\hat{p}_\xe) = 1$.

\section{Results}
\label{sec:results}

\subsection{NPRG fixed point solutions}

We present in the following the various fixed point solutions that we find within the NPRG approach.
As can be inferred from the $\beta$ functions (\ref{eq:betaflow}), all the set ($\hat{g} = 0, \gamma$) is a continuous line of fixed points. Within them, there are particular points where the flow along the $\gamma$ direction is vanishing even for a nonzero but small coupling $\hat{g}$. 
These fixed points are EW, EWA and EWU for which the coupling $\hat{g}_\kappa \rightarrow \hat{g}_* = 0$ (presented below). Quite generally, we find in this sector the same results as TF within the  perturbative approach.
In particular, we recover the same stability conditions as TF and find the EW exponents  $\chi_{\ew} = (2 - d)/2$ and $z_{\ew} = 2$ since $\eta_*^{\td} =  \eta_*^{\xa} =  \eta_*^{\xe} = 0$ for  these three weak-coupling solutions. 
It is further clear that  the isotropic fixed point solutions (with $\gamma_* = 1$) of the AKPZ equation are also  solutions of the KPZ equation. These solutions are therefore independent of the splitting into the sector dimensions $d_\xa$ and $d_\xe$, and depend only on the total dimension $d$.

Finally, let us  define for convenience the rescaled coupling constant $\tg' = v_d \tg / 4$ with  $v_d^{-1} = 2^{d-1} \pi^{d/2} \Gamma(d/2)$, that will be used for the graphical representation of the RG flows throughout this article.

\subsubsection{Isotropic Edwards-Wilkinson fixed point (EW)}
The isotropic EW fixed point corresponds to $(\hat{g}_*,\gamma_*) = (0,1)$.  This fixed point is repulsive (resp.\ attractive)
in the $\hat{g}$ direction for $d\le 2$ (resp.\ $d> 2$). In the direction of the anisotropy $\gamma$, the EW fixed point is attractive for $d < \sqrt{8}$ and repulsive for $d > \sqrt{8}$.

\subsubsection{Anisotropic Edwards-Wilkinson fixed point (EWA)}
The anisotropic EWA fixed point is located at $(\hat{g}_*,\gamma_*) = (0,(4 - d_\xa d) /(d d_\xe -4))$. It is repulsive (resp.\ attractive)
in the $\hat{g}$ direction for {\bf $d < 2$} (resp.\ $d> 2$). At exactly $d = 2$, EWA is repulsive (resp.\ attractive) for $d_\xa \geq 1 + 1/\sqrt{5}$ (resp.\ $d_\xa < 1 + 1/\sqrt{5}$ ) . In the direction of the anisotropy $\gamma$, the EWA fixed point is attractive for $d < - \Delta / 2 + \sqrt{(\Delta/2)^2 + 8}$ and $\sqrt{8} < d < \Delta / 2+ \sqrt{(\Delta/2)^2 + 8}$ and repulsive for $  - \Delta / 2 + \sqrt{(\Delta/2)^2 + 8} < d < \sqrt{8}$ and $d > \Delta / 2 + \sqrt{(\Delta/2)^2 + 8}$ with $\Delta = d_\xa - d_\xe$.

\subsubsection{Uniaxial Edwards-Wilkinson fixed point (EWU)}
The uniaxial EWU fixed point  at $(\hat{g}_*,\gamma_*) = (0,0)$ corresponds to the situation when $\lambda_\xe$ is zero.  It is repulsive (resp.\ attractive) in the $\hat{g}$ direction for $d < 2$ (resp.\ $d> 2$). Exactly at $d = 2$, EWU is repulsive (resp.\ attractive) in the $\hat{g}$ direction for $d _\xa \geq - 1 + \sqrt{5}$ (resp.\ $d _\xa < - 1 + \sqrt{5}$).
In the direction of the anisotropy $\gamma$, the EWU fixed point is repulsive for $d < - \Delta / 2 + \sqrt{(\Delta/2)^2 + 8}$ and attractive for larger values of $d$. A second EWU$_\infty$ fixed point at $(0,\infty)$ corresponds to the reverse situation when $\lambda_\xa$ is equal to zero.

\subsubsection{Isotropic Transition fixed point (T)}
The transition fixed point T is located at $\gamma_{*}  = 1$ and $\hat{g}_{*} > 0$ and exists for $d > 2$. It is always repulsive in the $\hat{g}$ direction. In the direction of the anisotropy $\gamma$ it is stable against anisotropic perturbations for small $d$ but becomes unstable for larger values of $d$ when the TA fixed point crosses it. The exponents are exactly known:  $\chi_T = 0$ and $z_T = 2$ in all dimensions. Within our approximation scheme, they are numerically found to be slightly negative (e.g.\ in $d = 3$ we find $\chi_T = -0.08$ (resp.\ $-0.12$) in NLO (resp.\ LPA') approximation) which physically implies $\chi_T=0$.   

\subsubsection{Anisotropic Transition fixed point (TA)}
The anisotropic transition fixed point is unstable and located at $\hat{g}_{*} > 0$. It is found for $d > 2$ and when the difference between the sector dimensions is sufficiently large so that the anisotropic fixed point A is present. 
TA is always repulsive in the $\hat{g}$ direction. 
In the direction of the anisotropy, the TA fixed point is attractive for small values of $d$ when it is in the $\gamma < 0$ quadrant but becomes unstable when it crosses the TU fixed point and moves into the $0 < \gamma_* < 1$ quadrant.
TA reverts to attractive when it crosses the T fixed point at $\gamma_* = 1$ and turns into the $\gamma_* > 1$ quadrant.
For $d > 2$ and small differences between the sector dimensions, TA merges with the TU fixed point at $\gamma_{*} = 0$, see Fig.\ \ref{fig:3d-flow}.

\subsubsection{Uniaxial Transition fixed point (TU)}
The uniaxial transition fixed point is unstable and located at $\hat{g}_{*} > 0$ and $\gamma_{*}  = 0$. It is found for $d > 2$ when the difference between the sector dimensions is sufficiently large so that the uniaxial  U fixed point is present.
TU is always unstable in the $\hat{g}$ direction. In the direction of the anisotropy, TU is attractive for small values of $d$ when the TA fixed point is in the $\gamma < 0$ quadrant but becomes unstable when TA crosses it to move into the $\gamma > 0$ quadrant.
For $d > 2$ and a decreasing difference between the sector dimensions,  TU merges with the TA and the A fixed point before it annihilates with U, see Fig.\ \ref{fig:3d-flow}.

\subsubsection{Isotropic Strong Coupling fixed point (SC)}
The strong coupling fixed point SC is located at $\gamma_{*}  = 1$ and $\hat{g}_{*} > 0$ and describes the isotropic rough phase of the KPZ equation. 
Within the NLO approximation, the associated exponents are in  good agreement with the numerical ones in $d=2$ and $d=3$  \cite{Canet10,Canet11a,*Canet12Err,Kloss12}. The quality of the NLO approximation in the isotropic case 
 deteriorates with increasing dimension and it does not yield reliable quantitative results above typically $d \simeq3.5$. 
In all dimensions at the LPA', and in all dimensions $d \lesssim 3.5$ at the NLO approximation, we find the SC fixed point to be locally stable and fully attractive.

\subsubsection{Anisotropic  fixed point (A)}
The anisotropic   fixed point A is located at $\gamma_{*}  < 0$ and $\hat{g}_{*} > 0$. It is fully attractive for $d < 2$ and for $d = 2$ when $d_\xa > 1 + 1 / \sqrt{5}$. For $d = 2$ and $d_\xa = 1 + 1 / \sqrt{5}$ it coincides with the EWA fixed point and moves to the negative quadrant for smaller values of $d_\xa$.
For $d > 2$, A is found for a large enough splitting between the sector dimensions. 
In that case it is also fully attractive but the dominant exponents around A may become complex.
To the best of our knowledge, this fixed point was not found yet and may correspond to a new rough anisotropic phase. 
Unfortunately, we find it only at non-integer dimensions. In consequence, its physical role  is unclear.

\subsubsection{Uniaxial fixed point (U)}
The uniaxial fixed point U is located at $\gamma_* = 0$ and $\hat{g}_* > 0$. It is present  for $d < 2$ and in $d = 2$ when $d_\xa  > -1 + \sqrt{5}$. U is attractive in the direction $\hat{g}$, but repulsive in the anisotropy direction $\gamma$. At $d = 2$ and $d_\xa  = -1 + \sqrt{5}$ it crosses the EWU fixed point and moves to the unphysical quadrant for smaller values of $d_\xa$. For $d > 2$ and a decreasing difference between both sector dimensions, U annihilates with the transition fixed point TU.

\subsection{Phase diagram}
\label{sec:phasediagram}
After the characterization of the various fixed point solutions, let us discuss the phase diagram of the AKPZ equation in the $(\hat{g}',\gamma)$ plane. 
For EW, EWU and EWA we recover the same stability conditions as TF in Ref.\ \cite{taeuber02}.
That is, the stability of the weak-coupling fixed points against anisotropic perturbations is changed by the EWA fixed point, which moves along the $(\hat{g}=0,\gamma)$ axis as a function of $d_\xa$ and $d_\xe$, and interchanges its stability  with the other fixed points EWU, EW and EWU$_{\infty}$ when crossing them. The sector dimensions $d_\xa$ and $d_\xe$ we find at which these crossings occur agree with those found by TF.
Apart from this  stability condition in the $\hat g$ direction reflecting the effect of anisotropic perturbations,   stability changes  also occur in the $\gamma$ direction to EW, EWU and EWA. This effect is already visible at the perturbative level, but requires nonperturbative methods to be thoroughly studied, since the crossing fixed points can become strong coupling in certain parameter regimes as detailed in the appendix D.
This analysis goes beyond the one of TF. In the following we will show how the
inclusion of the strong-coupling part changes qualitatively our understanding of the phase diagram.

\subsubsection{The case $d < 2$}
\begin{figure}[tbh]
  \centering
\includegraphics[width=80mm]{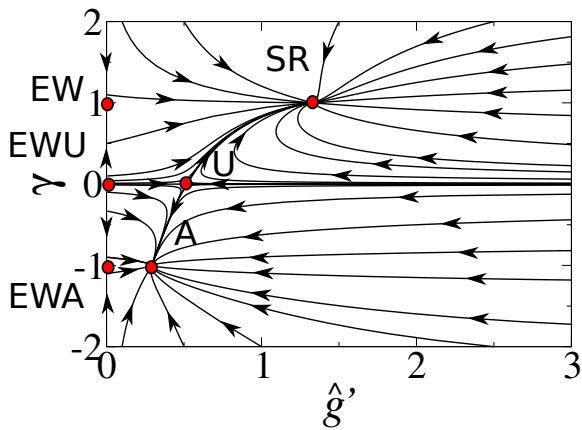}
  \vspace{-4mm}
  \caption{%
(Color online) Typical RG trajectories in the $(\tg' _\kappa ,\gamma _\kappa )$ plane for $d < 2$ (here $d_\xa = d_\xe = 0.8$). All weak coupling fixed points EW (with $\gamma _* = 1$), EWU (with $\gamma _* = 0$) and EWA (with $\gamma _* = (4 - d_\xa d) /(d d_\xe -4) $) are unstable in the $\hat g$ direction and we find two locally fully stable strong-coupling fixed points SC (with $\gamma _* = 1$) and A (with $\gamma _* < 0$). Moreover, there is another U fixed point with $\gamma _* = 0$  which is attractive along the $\gamma=0$ axis and repulsive in the $\gamma$ direction}.
  \label{fig:smalld-flow}
\end{figure}

To begin, the situation $d$ below 2 is depicted in Fig.\ \ref{fig:smalld-flow}.  All the weak-coupling fixed points EW, EWU and EWA are unstable in the $\hat g$ direction and three additional fixed points are found; one fixed point U is unstable in the $\gamma$ direction while attractive in the $\hat g$ direction along the $\gamma=0$ axis, and two locally  fully attractive SC and A fixed points. When both nonlinear couplings have the same sign, which corresponds to the flow in the positive $\gamma$ quadrant of Fig.\ \ref{fig:smalld-flow}, the flow is always attracted by the SC fixed point and the well known  isotropic KPZ rough phase is found.
On the other hand, if the couplings have opposite signs, which corresponds to the flow in the negative $\gamma$ quadrant, the flow is always driven towards the A fixed point. 

\subsubsection{The case $d = 2$ }
\begin{figure}[tbh]
  \centering
\includegraphics[width=80mm]{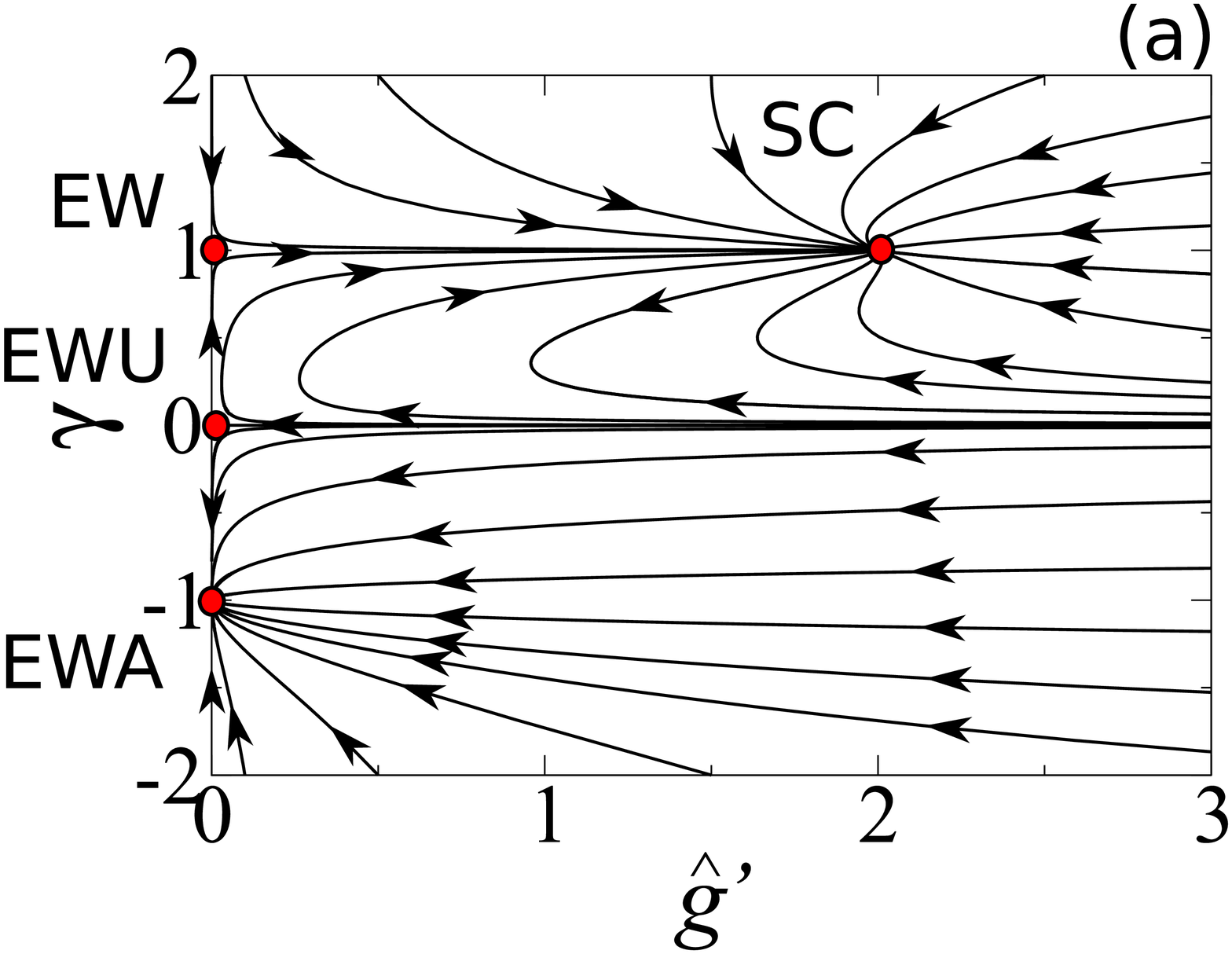}
\includegraphics[width=80mm]{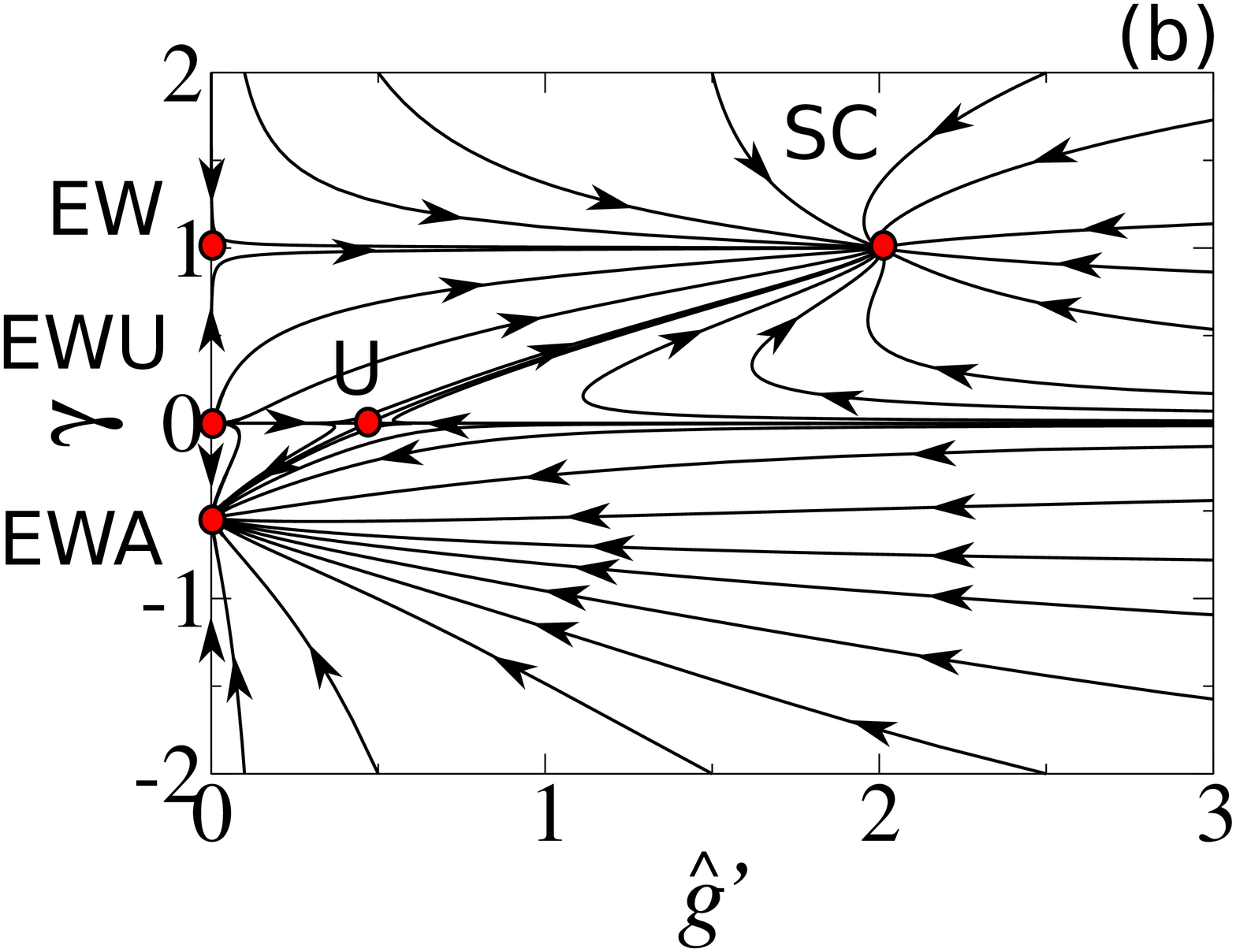}
\includegraphics[width=80mm]{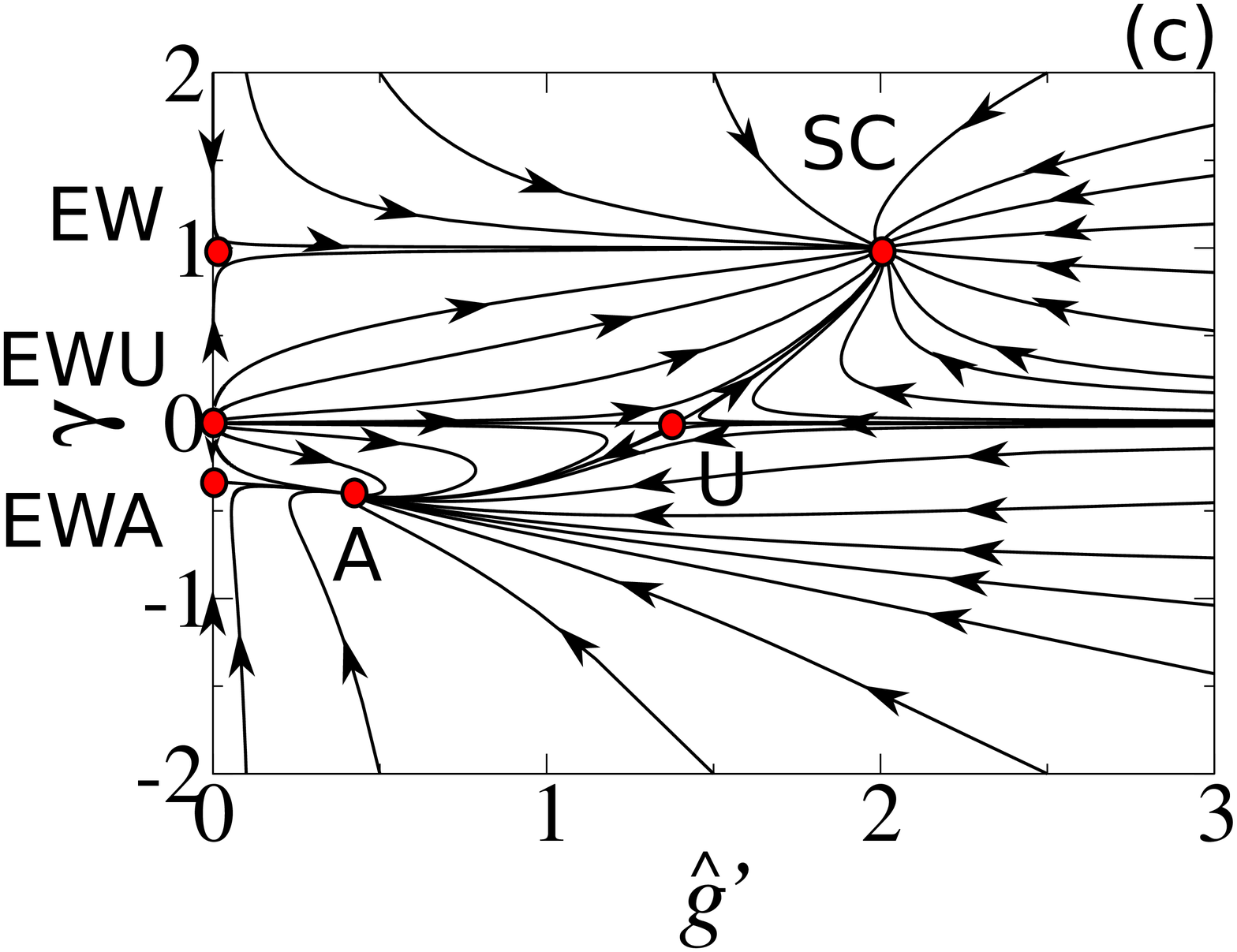}
  \vspace{-4mm}
  \caption{%
(Color online) RG	trajectories	in	the	$(\hat{g}'_\kappa,\gamma_\kappa)$	plane
at $d = 2$  with an increasing difference between the sector
dimensions (from up do down). Panel (a) represents a typical flow for $d_\xa < -1 + \sqrt{5}$ (here  $d_\xa = d_\xe = 1$, which corresponds to Wolf's result \cite{Wolf91}). Both the anisotropic fixed point EWA and the isotropic SC fixed point are fully attractive. The uniaxial fixed point EWU is attractive in the $\hat{g}$ direction but unstable in the anisotropy direction.  Panel (b): Typical flow for $ -1 + \sqrt {5} < d_\xa < 1 + 1/\sqrt {5} $ (here $d_\xa= 1.3$ and $d_\xe = 0.7$). EWU is now repulsive in the $\hat{g}$ direction. Panel (c): Typical flow for $d_\xa > 1 + 1/\sqrt  {5}$ (here $d_\xa= 1.5$ and $d_\xe = 0.5$). EWA becomes repulsive in the $\hat{g}$ direction because the anisotropic fixed point A crosses it. Note that in all cases the isotropic fixed point SC is not affected by the weak coupling stability changes and is always fully attractive. 
}
  \label{fig:2d-flow}
\end{figure}

 Exactly at $d = 2$ the  stability in the $\hat g$ direction of  the weak-coupling fixed points EW, EWU and EWA depends on the splitting in the sector dimensions. The RG flow for an increasing difference between $d_\xa$ and $d_\xe$ is depicted in Fig.\ \ref{fig:2d-flow}. 
 In this figure, the panel (a)  corresponds to Wolf's result \cite{Wolf91} with $d_\xa = d_\xe = 1$. That is, the EW fixed point is repulsive and EWU and EWA are both attractive in the $\hat g$ direction. When both nonlinear couplings have the same sign, the flow is always attracted by the SC fixed point. On the other hand, if the couplings have opposite signs, the flow is always driven towards the weak-coupling EWA fixed point. 
The situation changes when the difference between the sector dimensions increases. 
For  $d_\xa > -1 + \sqrt{5}$, first EWU becomes repulsive in the $\hat g$ direction, see Fig.\ \ref{fig:2d-flow} (b). This stability change is induced by the U fixed point, which crosses EWU at $d_\xa = -1 + \sqrt{5}$ and enters the physical quadrant.
Since U is unstable in the $\gamma$ direction the flow is still  attracted towards SC or EWA if $\gamma \neq 0$.
However, if the difference between the two sector dimensions is further increased above $d_\xa > 1 + 1/\sqrt{5}$,  EWA is also crossed by A and turns repulsive in the $\hat g$ direction, see Fig.\ \ref{fig:2d-flow} (c). For positive $\gamma$ the SC fixed point continues to be the only attractive fixed point, but in the negative $\gamma$ quadrant, the flow is driven towards A, similarly to the situation in $d < 2$. 

\subsubsection{The case $ 2 < d \leq 3$ }
\begin{figure}[tbh]
  \centering
\includegraphics[width=80mm]{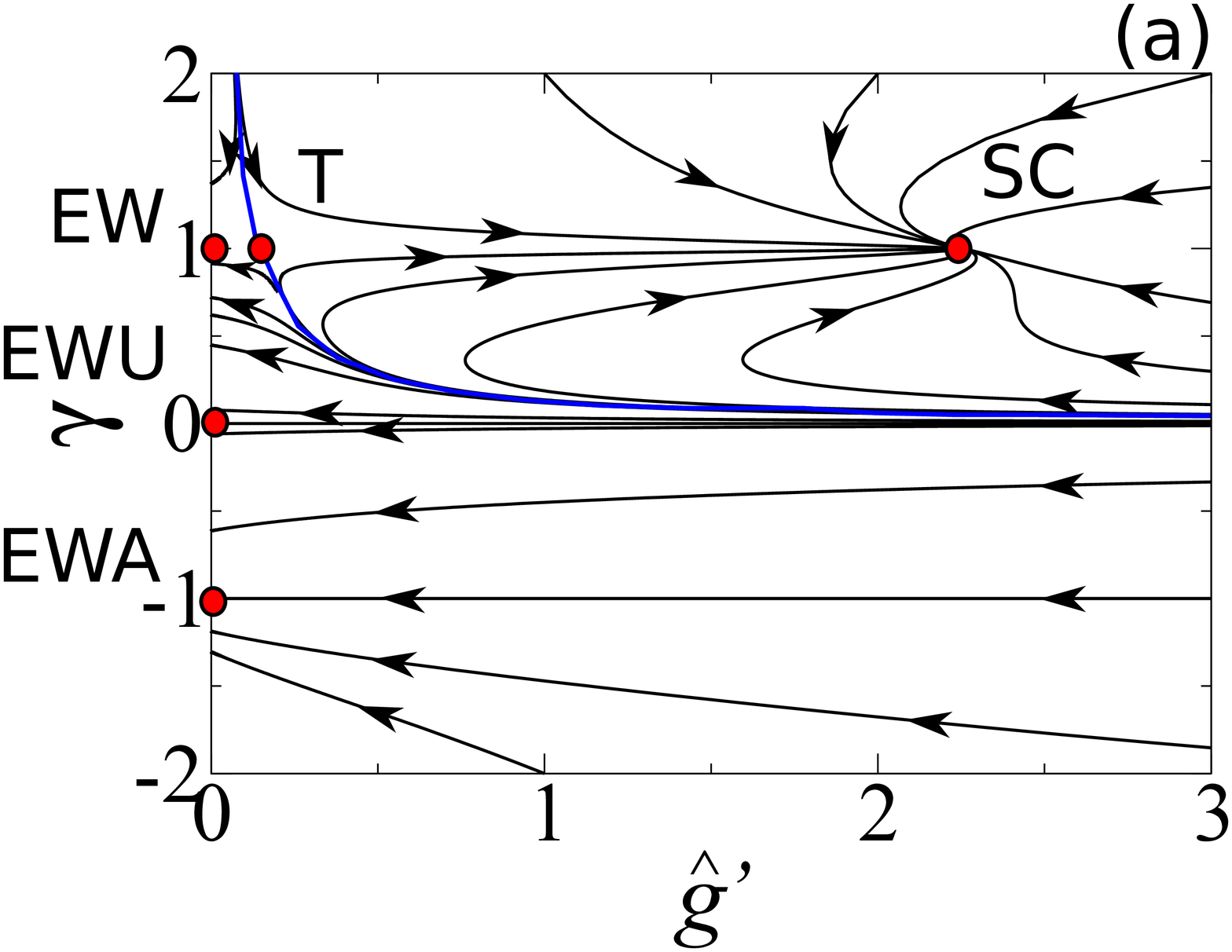}
\includegraphics[width=80mm]{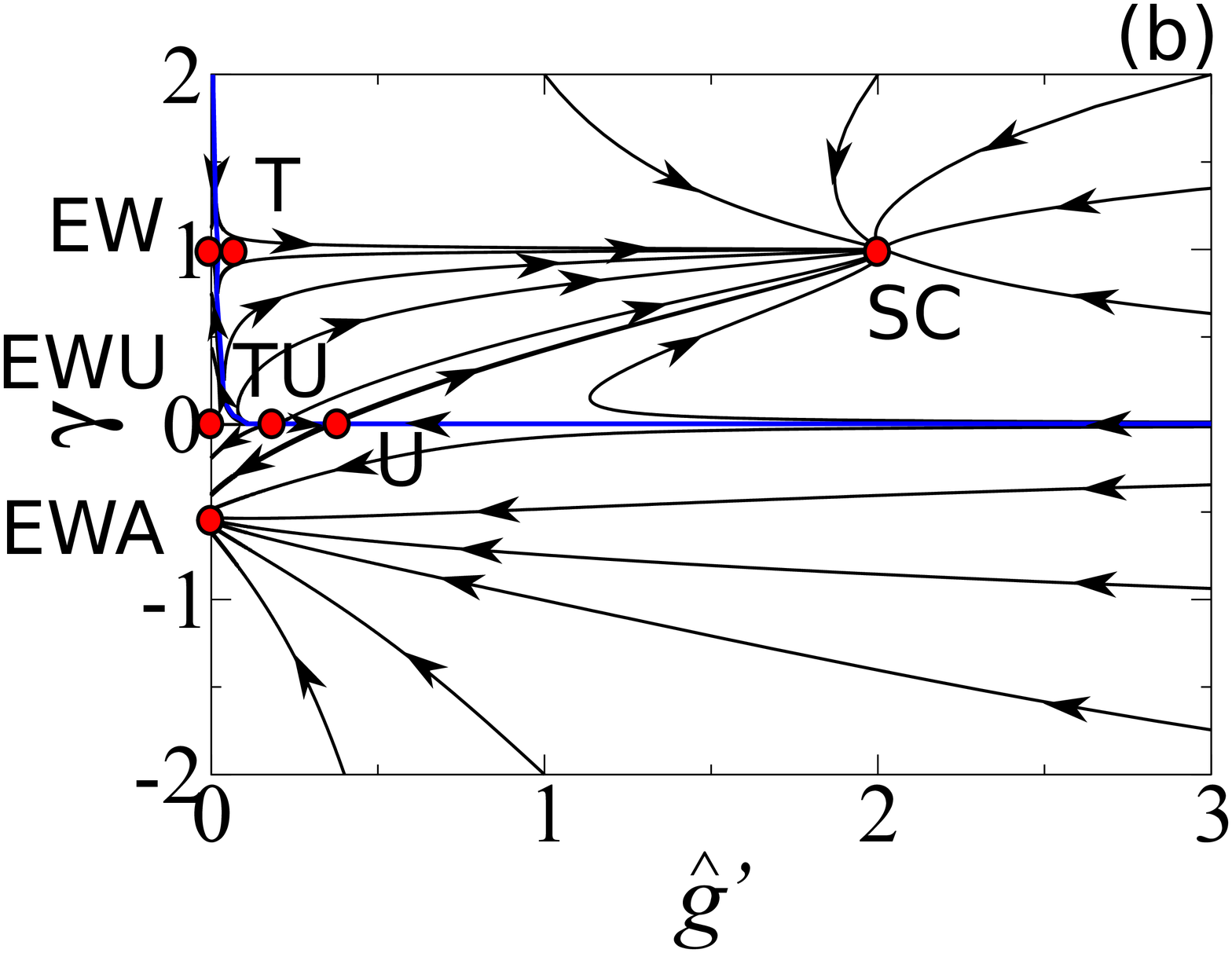}
\includegraphics[width=80mm]{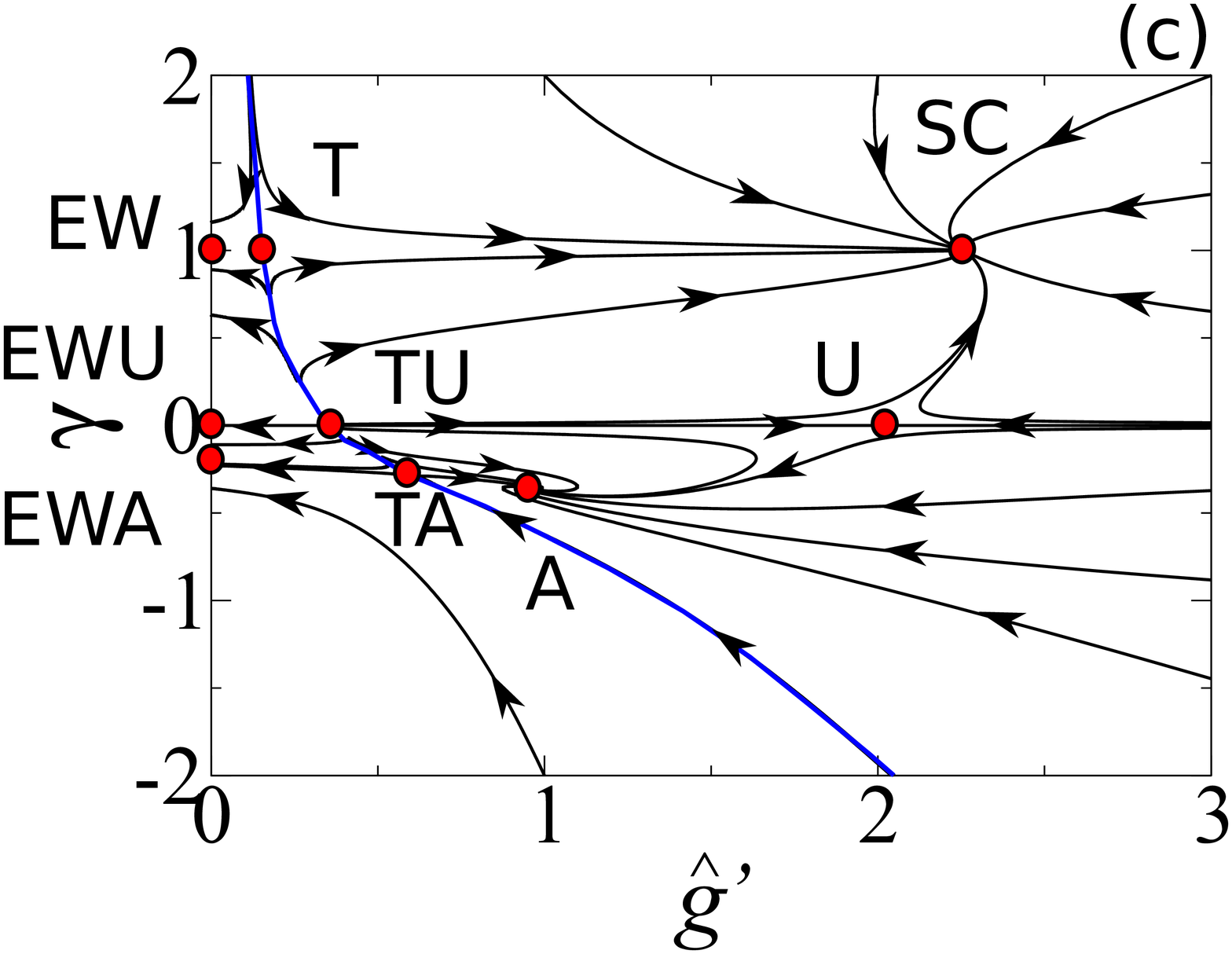}
  \vspace{-4mm}
  \caption{%
(Color online) RG	trajectories	in	the	$(\hat{g}'_\kappa,\gamma_\kappa)$	plane for $d > 2$ but with small $d - 2$. The difference between the sector dimensions increases along the panels. The three weak coupling fixed
points EW, EWU and EWA are attractive in the $\hat g$ direction in all panels.
Panel (a): Typical flow for$  d_\xa \lesssim 1.2 $ (here $d_\xa =  d_\xe = 1.1$). One unstable transition fixed point T ($\gamma _* = 1$) is present.
Panel (b): Typical flow for $ 1.2 \lesssim  d_\xa \lesssim 1.4 $ (here $d_\xa = 1.31$ and $d_\xe = 0.702$). Two unstable transition fixed points T and TU ($\gamma _* = 0$)  are present.
Panel (c): Typical flow for $ 1.4 \lesssim  d_\xa $ (here $d_\xa = 1.7$ and $d_\xe = 0.5$).
Three unstable transition fixed points T, TU and TA ($\gamma _* < 0$) are present. 
}
  \label{fig:22d-flow}
\end{figure}

For $d > 2$ all three fixed points EW, EWU and EWA become attractive in the $\hat g$ direction. More precisely, among these fixed points, the ones that are    attractive in the $\hat g$ direction in $d = 2$  remain so in $d>2$, 
 whereas for the others that are repulsive in $d = 2$, they are crossed in $d>2$ by a transition fixed point  as shown in Fig.\ \ref{fig:22d-flow} and thereby  become stable in the $\hat g$   direction.

\begin{widetext}
 
\begin{figure}[tbh]
  \centering
 \begin{minipage}{8cm}
 \hspace{-6mm}
\includegraphics[width=80mm]{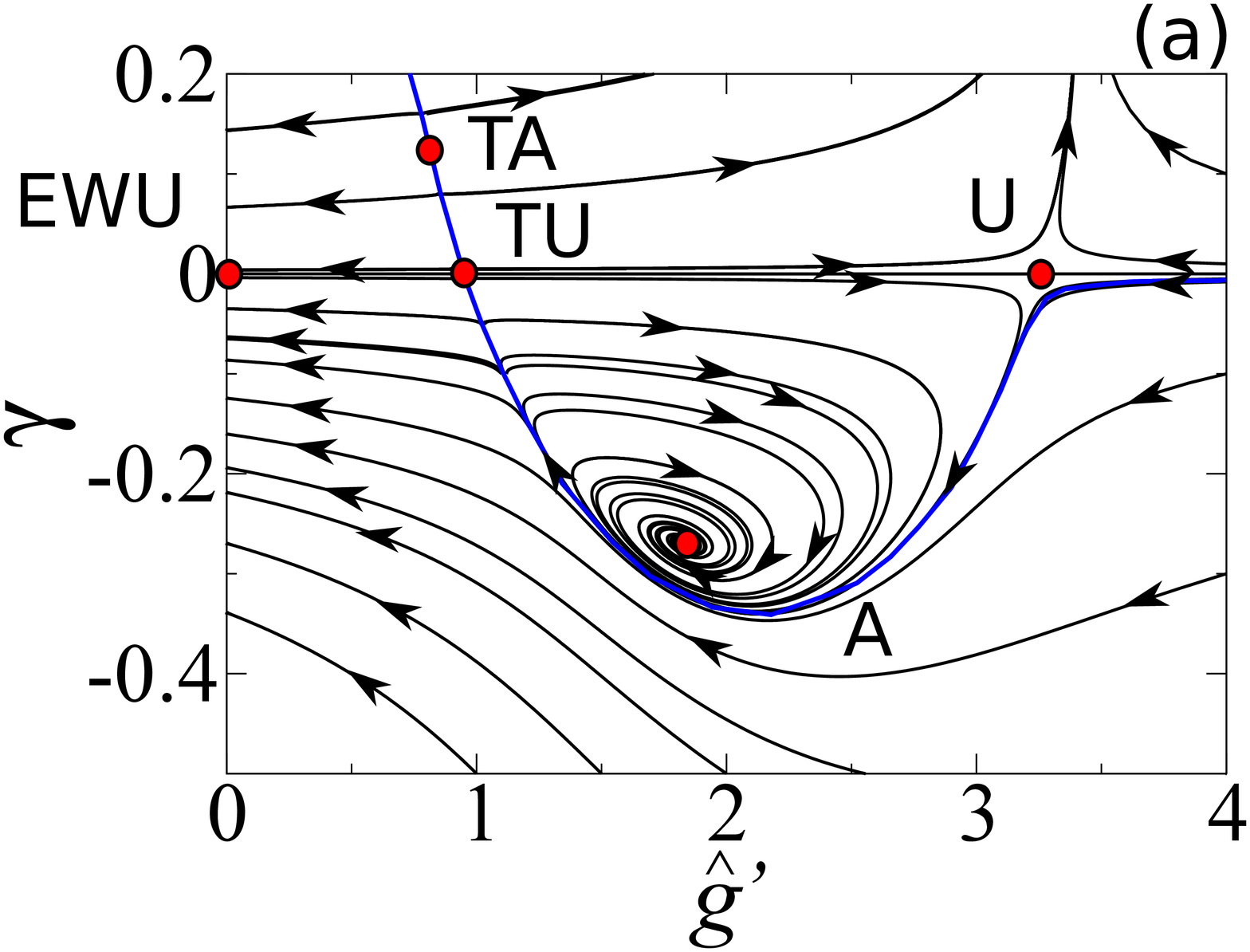}
\end{minipage}
 \begin{minipage}{8cm}
\includegraphics[width=80mm]{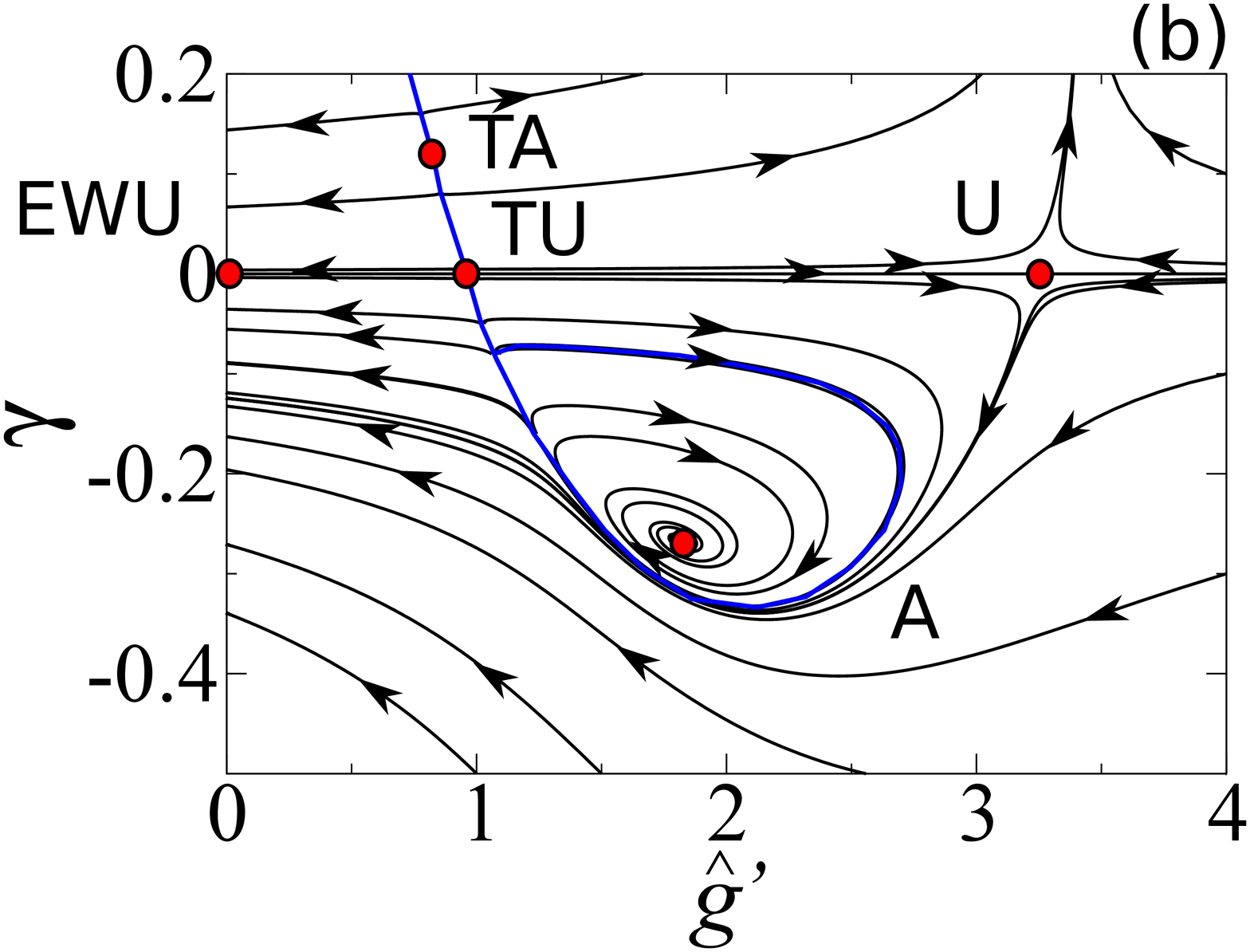}
\end{minipage}
 \begin{minipage}{8cm}
 \hspace{-6mm}
\includegraphics[width=80mm]{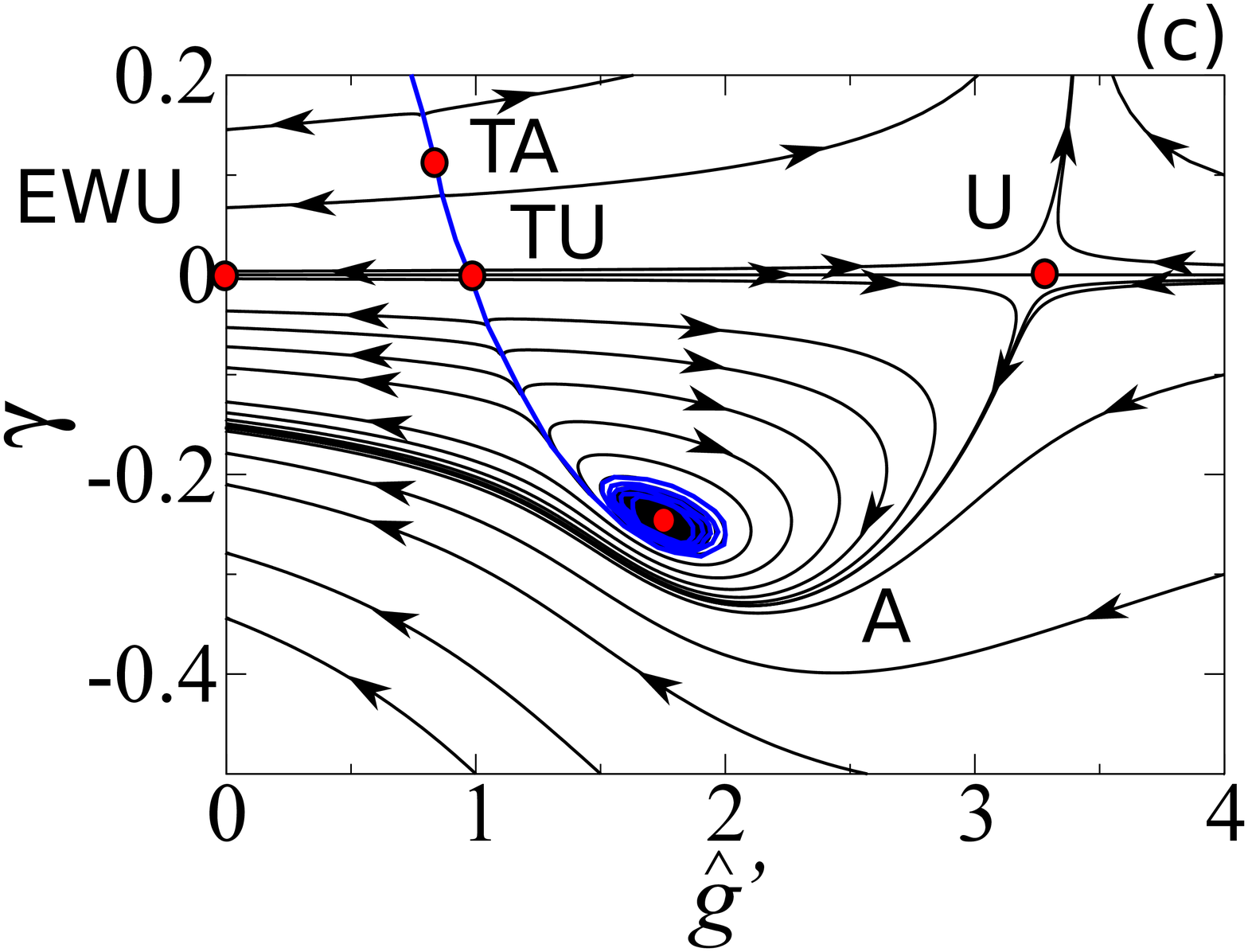}
\end{minipage}
 \begin{minipage}{8cm}
\includegraphics[width=80mm]{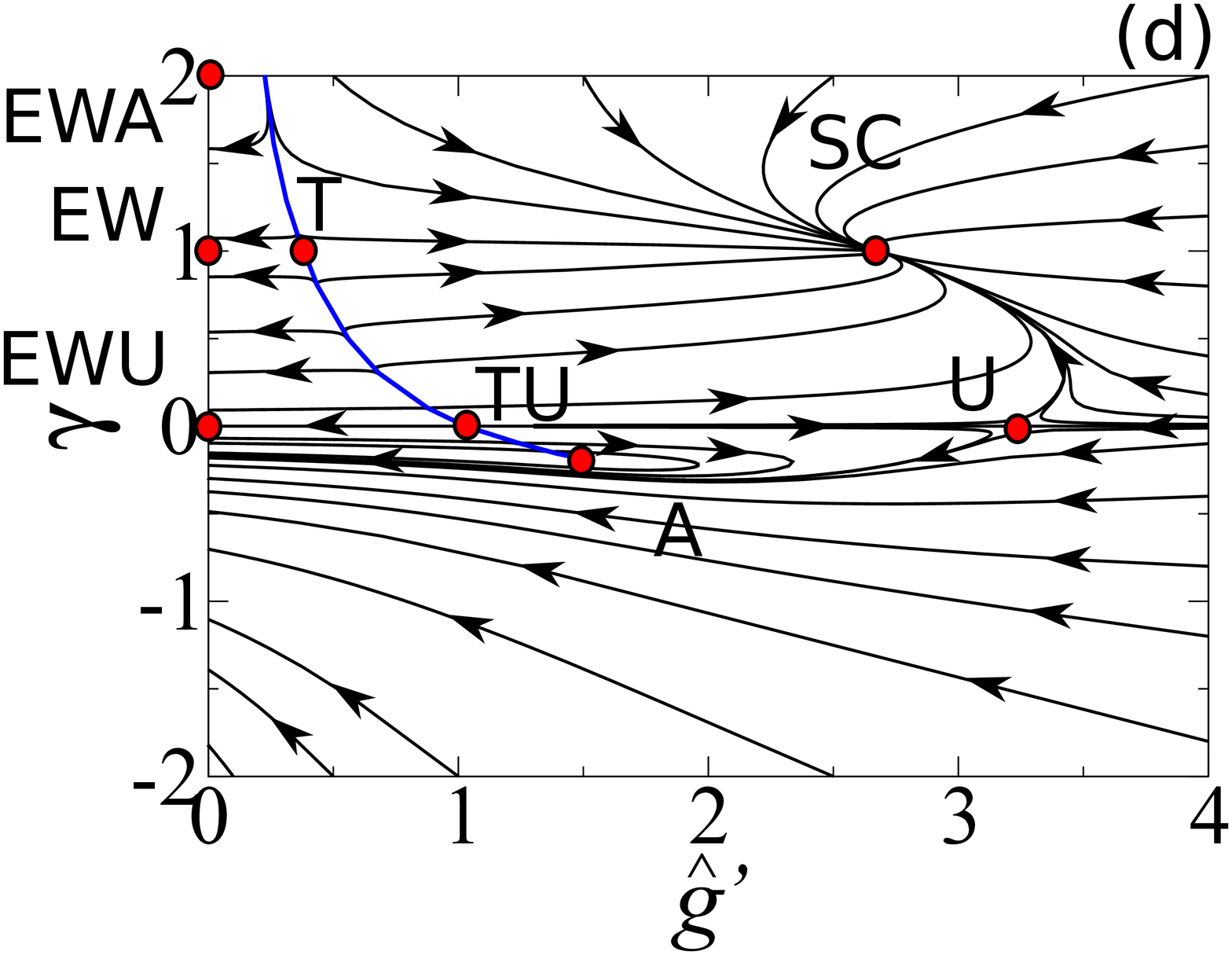}
\end{minipage}
 \begin{minipage}{8cm}
 \hspace{-6mm}
\includegraphics[width=80mm]{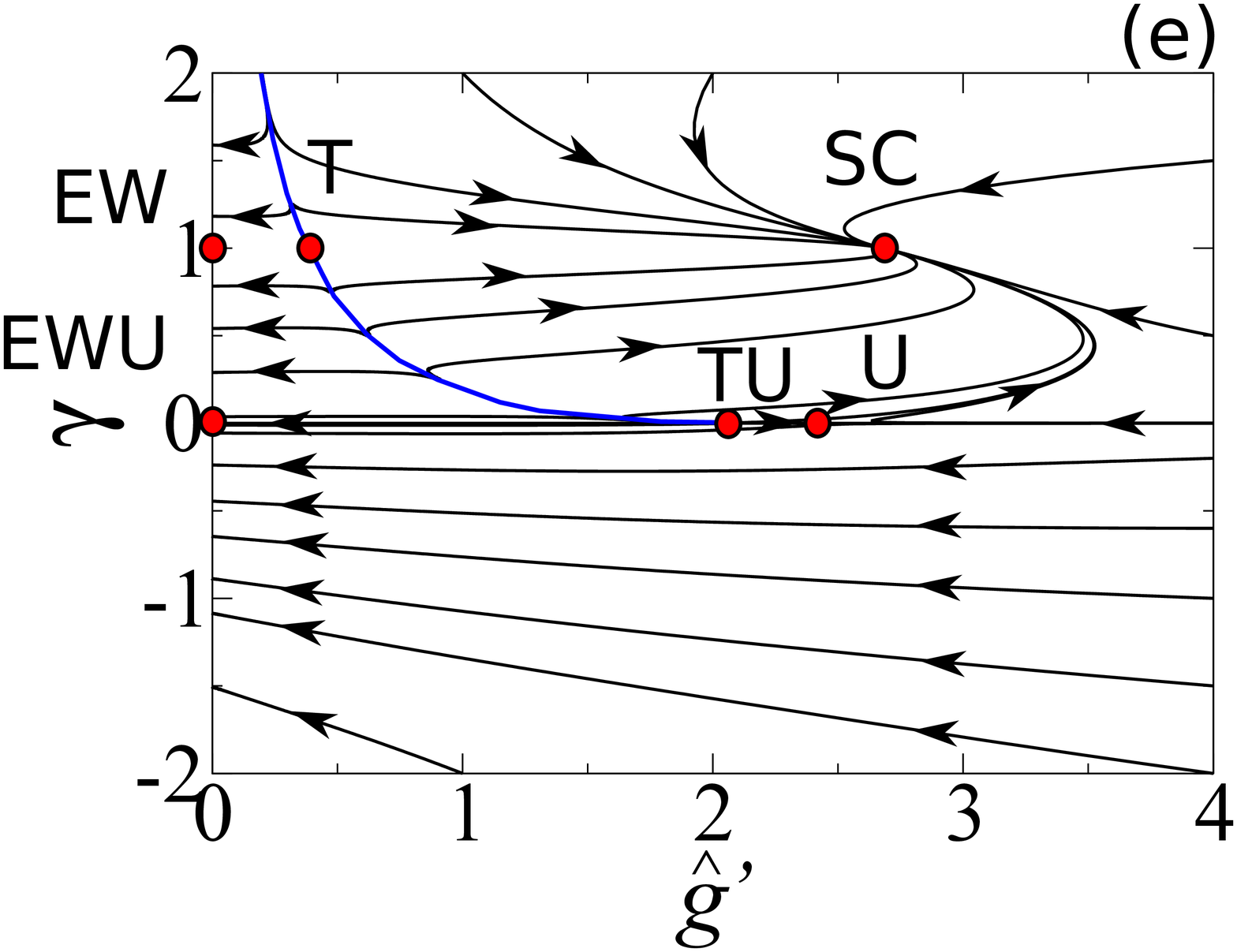}
\end{minipage}
 \begin{minipage}{8cm}
\includegraphics[width=80mm]{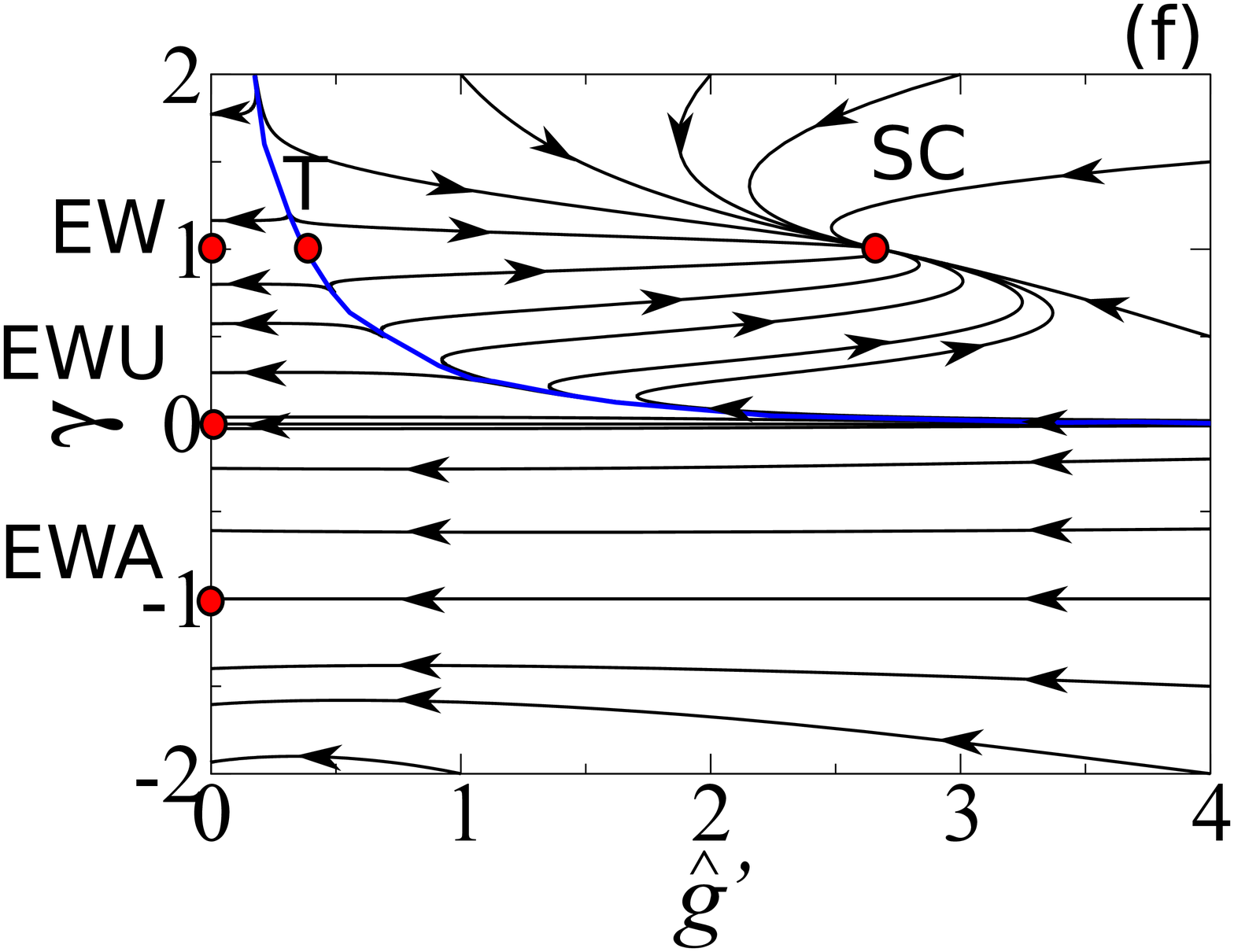}
\end{minipage}
  \vspace{-4mm}
  \caption{%
(Color online) RG trajectories in the $(\tg _\kappa ,\gamma _\kappa )$ plane for $d = 3$ and a decreasing difference between both sector dimensions (from up do down). Panel (a): The fixed point A is enclosed by the $\gamma = 0$ axis and the separatrix (highlighted in blue) and shows a spiral flow. (here $d_\xa = 2.04855$ and $d_\xe = 0.95145$). Panel (b): The separatrix has formed a closed loop around A. Inside this loop, the flow is directed towards A, outside and for $\gamma < 0$ the couplings  always flow to the weak coupling regime. (here $d_\xa = 2.047$ and $d_\xe = 0.953$). Panel (c): The enclosed area around A decreases in size. (here $d_\xa = 2.035$ and $d_\xe = 0.965$). Panel (d): The size of the attractive zone around A shrank down to zero. The TA fixed point has approached TU at $\gamma = 0$ and merges with it. (here $d_\xa = 2$ and $d_\xe = 1$). Panel (e): The endpoint of the separatrix where A was located approaches the TU fixed point and merges with it. TU and U approach each other. (here $d_\xa = 1.73$ and $d_\xe = 1.27$). Panel (f): TU and U have annihilated, so that the flow in the uniaxial case is always driven to the weak-coupling EWU. (here $d_\xa = d_\xe = 1.5$). 
}
  \label{fig:3d-flow}
\end{figure}

 \end{widetext}

When $d$  increases, the ordering in the $\gamma$ direction of the transition fixed points T, TU and TA may be different from the ordering of the  EW, EWU and EWA fixed points (see Figs.\ \ref{fig:22d-flow} and \ref{fig:23d-flow}). 
For $d$ larger than 2 and arbitrary splitting in the two sectors, we find the flow in certain cases to be more complex. For example in $d = 3$ we observe the flows displayed in Fig.\ \ref{fig:3d-flow}, which shows that for certain values of the splitting the dominant exponents around the A fixed point are complex. This gives a flow in form of spirals. Similar situations, with spiral flows around fixed points   were already observed for example in Ref.\  \cite{Delamotte08}.
Unfortunately, we do not find the A fixed point in $d = 3$ for integer values of the sector dimensions. If such a behavior had been present, it  would  imply a new anisotropic universality class in physical situations. 

\begin{figure}[tb]
  \centering
\includegraphics[width=80mm]{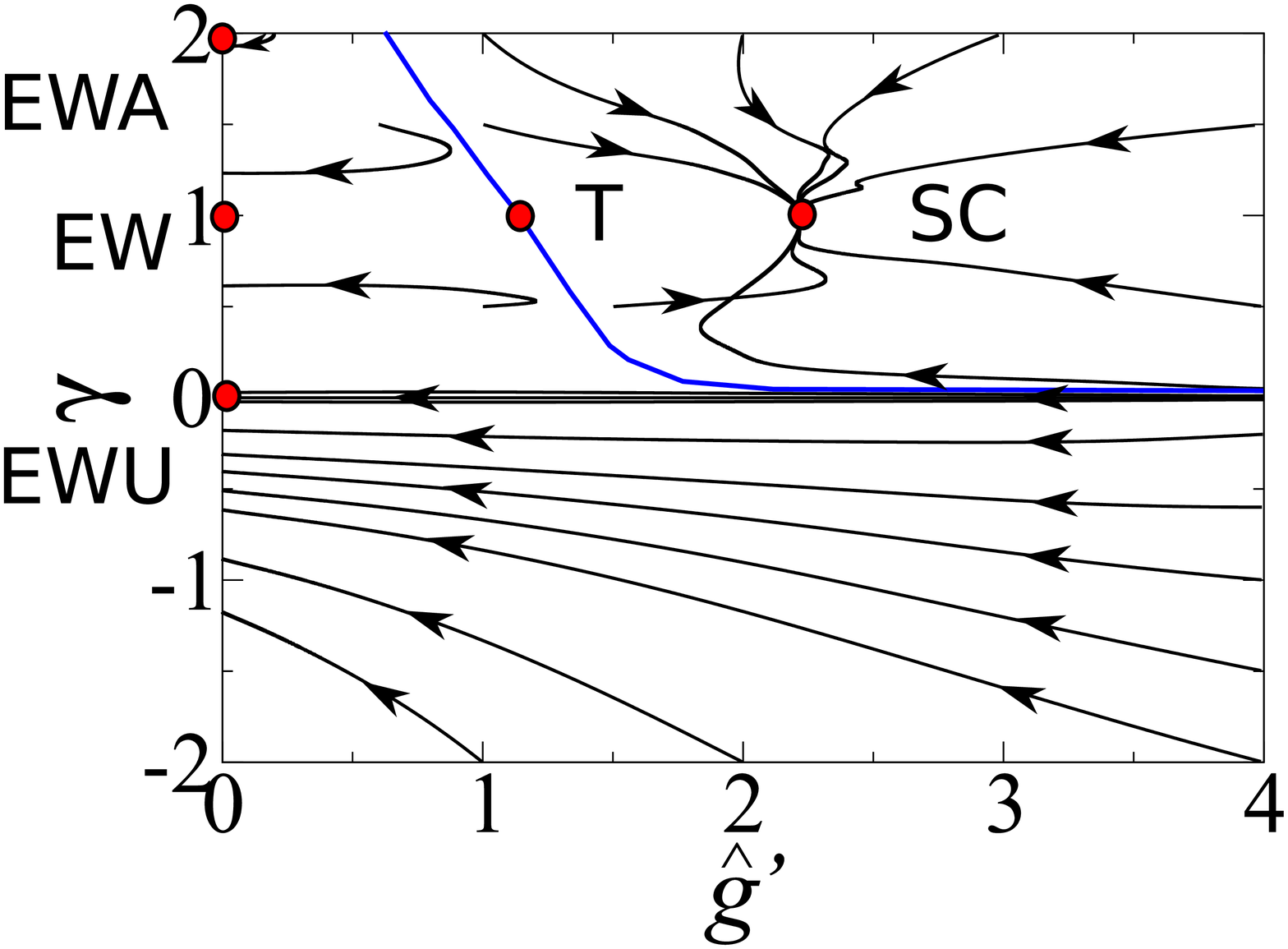}
  \vspace{-4mm}
  \caption{%
(Color online)  RG trajectories in the $(\tg' _\kappa ,\gamma _\kappa )$ plane for $d_\xa = 2$ and $d_\xe = 1$ in NLO approximation, to be compared with panel (d) of Fig.\ \ref{fig:3d-flow}. The strong coupling SC fixed point is locally fully attractive, but neither the A nor the U fixed points are present.
Note that the position of the separatrix (highlighted in blue) is only roughly estimated.
}
  \label{fig:nlo-flow}
\end{figure}

In the positive $\gamma$ region we did not find any case  in which the isotropic SC fixed point becomes locally unstable.
For the physically relevant case where both sector dimensions take integer values ($d_\xa =2, d_\xe =1$) we find the flow shown in Fig.\ \ref{fig:3d-flow}(d)  within the LPA' approximation. Within the NLO approximation, the flow diagram is slightly different, since the A and U fixed point are no longer present, as shown in Fig.\ \ref{fig:nlo-flow}. However, this does not affect the associated  physics. Indeed, within both approximations there is a critical coupling value  $\hat{g}_c$ in the positive $\gamma$ quadrant which is given by the separatrix (highlighted in blue) that goes through T. If the coupling at the initial scale is below this critical value, the flow renormalizes the coupling to zero and EW physics is obtained.  For a sufficiently large initial coupling $\hat{g} > \hat{g}_c$ the flow is attracted towards the SC fixed point, yielding the rough isotropic phase. 
At this point, it is not clear why the U and A fixed points are not present  for $d_\xa = 2$ and $d_\xe = 1$ in the NLO approximation, although they are found for $d < 2$ in both   LPA'  and NLO approximations. One possibility is that both fixed points A and U exist for more extreme splittings in the sector dimensions.  Unfortunately, this regime is numerically hard to explore within the NLO approximation because the double integral Eq.\  (\ref{eq:radialint}) becomes badly conditioned. 

\subsubsection{The case $ d > 3$ }

\begin{figure}[tbh]
  \centering
\includegraphics[width=80mm]{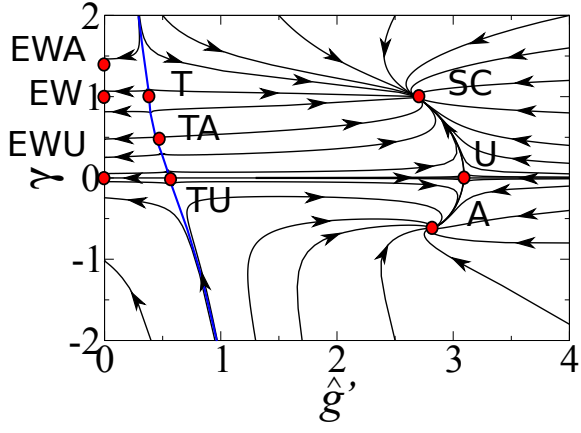}
  \vspace{-4mm}
  \caption{%
(Color online) RG trajectories in the $(\tg' _\kappa ,\gamma _\kappa )$ plane. Typical flow for $d$ approaching 3 with $d_\xe \sim 0.5$ (here $d_\xa = 2.5$ and $d_\xe = 0.5$). The unstable transition fixed points T ($\gamma _* = 1$), TA ($ 0 < \gamma _* < 1$) and TU ($\gamma _* = 0$) are present.  The ordering of the three transition fixed points along the $\gamma $ axis is no longer  the same as the one  of the weak coupling solutions (EWA, EW and EWU). It is useful to compare with Fig.\ \ref{fig:22d-flow}. 
 Note that we do not find a splitting in $d = 3$ which corresponds to scenario C in TF \cite{taeuber02} where the ordering becomes (EWA, EW and EWU) and (TA, T and TU).
}
  \label{fig:23d-flow}
\end{figure}

As already pointed out, the NLO approximation turns out to become quantitatively unreliable  in the isotropic strong coupling regime for  dimensions $d \gtrsim 3.5$ \cite{Canet10, Kloss12,Kloss14}. In large dimensions one typically do find the strong coupling fixed point but the associated critical exponents largely depend on the regulator. However, the quality of the NLO approximation  around the anisotropic or the uniaxial fixed point is \textit{a priori} not known.
In the physically relevant cases $d_\xa = d_\xe = 2$ and $d_\xa = 3 $ with $ d_\xe = 1$ we find the isotropic strong coupling fixed point always locally fully stable, both within the LPA' and the NLO approximations \footnote{We use $\alpha = 10$ in $d = 4$, compare Ref.\ \cite{Kloss14}}.
The flow is qualitatively similar to Fig.\ \ref{fig:3d-flow} (e) when $d_\xa = d_\xe = 2$, whereas it resembles  Fig.\ \ref{fig:23d-flow} when $d_\xa = 3 $ and $ d_\xe = 1$ \footnote{In both cases ($d_\xa = 3$ and $ d_\xe = 1$ ) and ($d_\xa = d_\xe = 2$ ) the fixed point TA ($\gamma_* > 1$) has crossed T ($\gamma_* = 1$) so that T and TA have interchanged their stability in the $\gamma$ direction. In contrast, for $d = 3$ we find $\gamma_* < 1$ for TA. For the flow diagram this is however of minor importance since both fixed points T and TA are unstable so that the RG trajectories in ($d_\xa = 3$ and $ d_\xe = 1$) are almost similar to the situation ($d_\xa = 2.5$ and $ d_\xe = 0.5$) and ($d_\xa = d_\xe= 2$) is almost similar to ($d_\xa = 1.73$ and $ d_\xe = 1.27$).}. 
In contrast to the former case in $d = 3$, we find the A fixed point for the integer splitting $d_\xa = 3$ and $d_\xe = 1$ at LPA', but it is not present at NLO. This indicates that the  LPA' (or even NLO) approximation may be unreliable in this regime of the flow.

\subsection{On the stability of the  isotropic rough phase}
\label{sec:stability}

In the previous section we stressed that the strong coupling behavior in the positive $\gamma$ quadrant is always driven by a unique and locally fully attractive SC fixed point. For non-integer dimensions we find other  fixed points A and U, but they lie in the negative $\gamma$ quadrant (resp.\ at $\gamma = 0$). As already mentioned, we never observe that these fixed points enter the positive $\gamma$ quadrant. Hence, they do not cross the SC fixed point and leave its stability unchanged. 
Physically, this implies that the isotropic rough phase is locally stable against anisotropic perturbations even for arbitrary non-integer sector dimensions.
Our work generalizes the one of TF \cite{taeuber02} since we are able to study the strong coupling regime. In agreement with TF we find that the isotropic KPZ equation may become unstable against anisotropic perturbations in the weak-coupling regime.
However, our results indicate that it is not possible to na{\"i}vely extrapolate  the perturbative results to the strong coupling regime in order to predict the stability of the SC phase. In contrast to the weak-coupling regime the isotropic strong coupling fixed point SC remains always locally fully stable.

\section{Conclusion}
\label{sec:conclusion}

In this manuscript, we have presented a NPRG analysis of the AKPZ equation.
We have derived the associated flow equations and solved them numerically to study the RG flow and their fixed point solutions. Two different approximations were applied throughout this study, the simpler LPA' approximation to obtain the qualitative features of the phase diagram and the more accurate NLO approximation in order to check our findings.

At weak coupling the LPA' approximation leads to the same equations (and stability conditions) as the perturbative ones in the study by TF, but moreover, with the NPRG approach we are able to study the strong coupling regime. We find that the isotropic rough KPZ phase is always locally stable against anisotropic perturbations. 
 The AKPZ equation provides no hint for the existence of an upper critical dimension of the strong-coupling phase (similarly to the conclusion drawn from the analysis of the KPZ equation with long-range noise \cite{Kloss14}).

In fact, if both nonlinear couplings have the same sign, the critical behavior of the AKPZ equation resembles much that of the standard KPZ equation and the critical exponents $\chi$ and $z$ only depend  on the total spatial dimension $d$ but not on the splitting in the two sector dimensions. That is, for $d \leq 2$ the flow is always driven towards the isotropic rough strong coupling fixed point. For $d > 2$ there is a critical coupling $\hat{g}_c$ below which the flow renormalizes the coupling to zero and EW physics is obtained.  In contrast, for sufficiently large coupling $\hat{g} > \hat{g}_c$ the flow is attracted towards the SC fixed point, describing the  isotropic rough phase. The anisotropy plays no role for  nonlinearities of same signs.

However, if the nonlinearities have different signs, the anisotropy  becomes relevant. 
The case $d_\xa = d_\xe = 1$,  initially studied by Wolf finding only weak coupling behaviors, turns out to be only a special case, as already pointed out by TF.
In the general situation, we find a new locally attractive anisotropic strong-coupling fixed point,  that could correspond to a new universality class. 
However, it is only present  for certain non-integer values of $d_\xa$ and $d_\xe$, and does not exist at integer sector dimensions. 

We hope that our results can have some relevance for the physics of cold atoms, considering the recently established connection with an anisotropic and periodic variant of the KPZ equation \cite{Chen13,Altman14}. The  AKPZ equation readily incorporates the effect of anisotropy but ignores that of periodicity. At the current stage it is however not possible to predict the importance of this missing ingredient and how it might change the physics. This work is left for further studies.

\section{Acknowledgments}
LC and TK thank the Universidad de la Rep\'ublica, NW the LPMMC and TK the Goethe Universit\"at Frankfurt and the Group of P.\  Kopietz for hospitality. 
This work has received financial support from PEDECIBA program, the ECOS France-Uruguay collaboration program and TK acknowledges financial support from the IIP in Natal. 
Numerical calculations where performed on the Computer System of High Performance (IIP and UFRN, Natal) and on HOUSEWIVES (Goethe Universit\"at, Frankfurt).
The authors would also like to thank B.\ Delamotte for stimulating discussions.

\begin{appendix}

\renewcommand{\theequation}{A\arabic{equation}}
\section*{Appendix A: perturbative analysis in the weak-coupling limit}
\setcounter{equation}{0}

In the limit $\tg_\kappa \rightarrow 0$, the anomalous dimensions simplify to
\begin{equation}  
 \etad_\kappa =  - \hat{I}^{\td \xn}_\kappa , \quad \eta^\xa_\kappa =  - \hat{I}^{\xa \xn}_\kappa , \quad \eta^\xe_\kappa =  - \hat{I}^{\xe \xn}_\kappa .
 \label{eq:eta0}
\end{equation}  
As emphasized in previous studies \cite{Kloss14}, the NLO approximation reduces to the LPA' approximation in that limit because the difference of the flowing functions $f^\xx$ to one is already of order $\tg$.
We can therefore take the limit of small coupling directly in the LPA' approximation Eq.\ (\ref{eq:simple}) and  calculate explicitly the
loop integrals Eqs.\ (\ref{eq:etaint}) to obtain
\begin{subequations}
  \begin{align}
  \etad_\kappa 
  &= \tg_\kappa [ d_\xa (d_\xa + 2) + d_\xe (d_\xe +2) \gamma_\kappa^2 + 2 d_\xa d_\xe \gamma_\kappa] A(d) , 
   \\
 \eta^\xa_\kappa & = - \tg_\kappa [d (d_\xa + d_\xe \gamma_\kappa) - 4] A(d) ,
   \\
 \eta^\xe_\kappa & = - \tg_\kappa \gamma_\kappa  [d (d_\xa + d_\xe \gamma_\kappa)  - 4 \gamma_\kappa] A(d) .
  \end{align}
  \label{eq:eta1}
\end{subequations}
These expressions are, up to the inverse sign convention and the extra factor
\begin{equation}  
 A(d) =   - \frac{ v_d}{2d (d + 2)} \int_{0}^{\infty} \! \! d \tq \, \tq^{d-1} \frac{ \partial_{\tq^2} r(\tq^2)}{ (1 + r(\tq^2))^2} ,
\end{equation}  
similar to the one-loop results of TF \cite{taeuber02}.
The integral $A(d)$ is a positive constant which depends only on the dimension and on the cutoff function.
Note that for the special value $d = 2$ this integral is independent of the regulator and we find $A(2) = 1 / (64 \pi)$.
The factor $A(d)$ can further be absorbed by a simple redefinition of $\hat{g}_\kappa$. For all  the weak-coupling fixed points EW, EWU and EWA we therefore recover exactly the same stability conditions as TF.

\renewcommand{\theequation}{B\arabic{equation}}
\section*{Appendix B: Rescaling}
\setcounter{equation}{0}

As already noted in Ref.\ \cite{taeuber02} the AKPZ equation can be rescaled such that either $\nu_{\xa} = \nu_{\xe}$ or $|\lambda_{\xa}| = |\lambda_{\xe}|$. The first variant is derived explicitly in this appendix, since the resulting rescaled AKPZ action is the one studied  throughout this work.
Performing the substitution
\begin{subequations}
\begin{align}
 h(t,\vx) &= \left( \frac{D}{\nu_\xa}\right)^{1/2}\gamma^{d_\xe / 4}h'(t',\vx') , \\
 \tilde{h}(t,\vx) &= \left( \frac{\nu_\xa}{D}\right)^{1/2} \gamma^{d_\xe / 4} \tilde{h}'(t',\vx') , \\
\vx_{\xa} &=  \vx'_{\xa}, \quad \vx_{\xe} = \gamma^{-1/2} \vx'_{\xe} ,  \quad t = t' / \nu_{\xa}, 
\end{align}
\end{subequations} 
in the action Eq.\ (\ref{eq:action}) where $\gamma = \nu_{\xa} / \nu_{\xe}$, we obtain the rescaled AKPZ action (to simplify notations, the primes ' are omitted):
\begin{align}
& {\cal S}[h,\tih]  \! = \!\! \int_{\bx}  \!\Bigl\{ \tih(\bx)\Bigl(\p_t h(\bx) - \,\nabla^2_{\xa} h(\bx) -  \,\nabla^2_{\xe} h(\bx) \nonumber \\
& - 
\frac{\sqrt{g^\xa_b}}{2}\! \left({\nabla_{\xa}} h(\bx)\right)^2 - 
\frac{\sqrt{g^\xe_b}}{2} \! \left({\nabla_{\xe}} h(\bx)\right)^2\Bigr) -  (\tih(\bx))^2 \Bigr\} .
\label{eq:rescal1}
\end{align} 
where
\begin{subequations}
\begin{align}
  \sqrt{g^\xa_b}& \equiv  \left( \frac{\lambda_{\xa}^2 D \gamma^{d_\perp / 2}}{\nu_\xa^{3}}\right)^{1/2}, \\
 \sqrt{g^\xe_b} &\equiv \left( \frac{\lambda_{\xe}^2 D \gamma^{d_\xe / 2+2}}{\nu_\xa^{3}}\right)^{1/2}  .
\end{align}
\label{eq:effCoupl}
\end{subequations} 
The rescaling  therefore amounts to the substitution 
$ \lambda_{\xa} \rightarrow  \sqrt{g^\xa_b}$ and 
 $\lambda_{\xe} \rightarrow \sqrt{g^\xe_b}$  
in the original AKPZ action with all the other constants set to unity.

\renewcommand{\theequation}{C\arabic{equation}}
\section*{Appendix C: flow integrals for the anomalous dimensions}
\setcounter{equation}{0}

In the NLO approximation, the flow integrals with zero external frequency and momentum  are
\begin{widetext}
\begin{subequations}
\begin{align}
\tI^{\td\td}_\kappa &= - \tg_\kappa \frac{ v_{d_\xa}v_{d_\xe}}{2} \int_{0}^{\infty} \! \! \!  d \tq_\xa \, \tq^{d_\xa-1}_\xa \! \! \! \int_{0}^{\infty} \! \!  \! d \tq_\xe \,  \tq^{d_\xe -1}_\xe \, (\tq_\xa^2 + \gamma_\kappa \tq_\xe^2)^2 \frac{  r(\tq^2) \, \hat k_\kappa(\hat q_\xa,\hat q_\xe) }{\tf^\lambda_\kappa(\hat q_\xa,\hat q_\xe) ( \hat l_\kappa(\hat q_\xa,\hat q_\xe))^3} , \\
\tI^{\td\xa}_\kappa &= \tg_\kappa \frac{ 3 v_{d_\xa}v_{d_\xe}}{4} \int_{0}^{\infty} \! \! \!  d \tq_\xa \, \tq^{d_\xa-1}_\xa \! \! \! \int_{0}^{\infty} \! \!  \! d \tq_\xe \,  \tq^{d_\xe -1}_\xe \,  \tq_\xa^2  (\tq_\xa^2 + \gamma_\kappa \tq_\xe^2)^2  \frac{ r(\tq^2) \,(\hat k_\kappa(\hat q_\xa,\hat q_\xe))^2}{\tf^\lambda_\kappa(\hat q_\xa,\hat q_\xe) (\hat l_\kappa(\hat q_\xa,\hat q_\xe))^4} ,  \\
\tI^{\td\xe}_\kappa &= \tg_\kappa \frac{ 3 v_{d_\xa}v_{d_\xe}}{4} \int_{0}^{\infty} \! \! \!  d \tq_\xa \, \tq^{d_\xa-1}_\xa \! \! \! \int_{0}^{\infty} \! \!  \! d \tq_\xe \,  \tq^{d_\xe -1}_\xe \,  \tq_\xe^2  (\tq_\xa^2 + \gamma_\kappa \tq_\xe^2)^2  \frac{ r(\tq^2) \,(\hat k_\kappa(\hat q_\xa,\hat q_\xe))^2}{\tf^\lambda_\kappa(\hat q_\xa,\hat q_\xe) (\hat l_\kappa(\hat q_\xa,\hat q_\xe))^4} ,  \\
\tI^{\td\xn}_\kappa &= \tg_\kappa \frac{ v_{d_\xa}v_{d_\xe}}{2} \int_{0}^{\infty} \! \! \!  d \tq_\xa \, \tq^{d_\xa-1}_\xa \! \! \! \int_{0}^{\infty} \! \!  \! d \tq_\xe \,  \tq^{d_\xe -1}_\xe \,  \tq^2  (\tq_\xa^2 + \gamma_\kappa \tq_\xe^2)^2 \, \frac{ (\partial_{\tq^2} r(\tq^2)) \hat k_\kappa(\hat q_\xa,\hat q_\xe) }{\tf^\lambda_\kappa(\hat q_\xa,\hat q_\xe) (\hat l_\kappa(\hat q_\xa,\hat q_\xe))^4} 
   \Bigl[ 3 \tq^2 \hat  k_\kappa(\hat q_\xa,\hat q_\xe) - 2 \hat l_\kappa(\hat q_\xa,\hat q_\xe) \Bigr] ,  \\
\tI^{\xa \td}_\kappa &= \tg_\kappa \frac{v_{d_\xa}v_{d_\xe} }{4d_\xa} \int_{0}^{\infty} \! \! \!  d \tq_\xa \, \tq^{d_\xa-1}_\xa \! \! \! \int_{0}^{\infty} \! \!  \! d \tq_\xe \,  \tq^{d_\xe -1}_\xe \,  \frac{ r(\tq^2)}{ \tf^\lambda_\kappa(\hat q_\xa,\hat q_\xe) (\hat l_\kappa(\hat q_\xa,\hat q_\xe))^3} \Bigl[ 
2 (\gamma_\kappa-1) (\hat q_\xa^2 \hat q_\xe^2 / \hat q^2)   \tf^\lambda_\kappa(\hat q_\xa,\hat q_\xe)  \hat l_\kappa(\hat q_\xa,\hat q_\xe)   \nonumber \\
& \qquad + (\tq_\xa^2 + \gamma_\kappa \tq_\xe^2)  \Bigl(  \tf^\lambda_\kappa(\hat q_\xa,\hat q_\xe)  \hat q_\xa \partial_{\tq_\xa} \hat l_\kappa(\hat q_\xa,\hat q_\xe) - \hat l_\kappa(\hat q_\xa,\hat q_\xe) \hat q_\xa \partial_{\tq_\xa} \tf^\lambda_\kappa(\hat q_\xa,\hat q_\xe) -2 (\hat q_\xa^2 / \hat q^2) \tf^\lambda_\kappa(\hat q_\xa,\hat q_\xe)\hat l_\kappa(\hat q_\xa,\hat q_\xe)  \Bigr)   \Bigr] ,  \\
\tI^{\xa\xa}_\kappa &= -\tg_\kappa \frac{v_{d_\xa}v_{d_\xe} }{4d_\xa}   \!\! \int_{0}^{\infty} \! \! \!  d \tq_\xa \, \tq^{d_\xa-1}_\xa \! \! \! \int_{0}^{\infty} \! \!  \! d \tq_\xe \,  \tq^{d_\xe -1}_\xe \,   \frac{\tq^{2}_\xa r(\tq^2)}{\tf^\lambda_\kappa(\hat q_\xa,\hat q_\xe) (\hat l_\kappa(\hat q_\xa,\hat q_\xe))^3} \Bigl[ 
- \Bigl(2 \hat q_\xa^2 -d_\xa (\tq_\xa^2 + \gamma_\kappa \tq_\xe^2) \Bigr) \tf^\lambda_\kappa(\hat q_\xa,\hat q_\xe)  \hat k_\kappa(\hat q_\xa,\hat q_\xe) \nonumber \\ & 
\qquad + (\tq_\xa^2 + \gamma_\kappa \tq_\xe^2) \Bigl(\tf^\lambda_\kappa(\hat q_\xa,\hat q_\xe) \tq_\xa \partial_{\tq_\xa} \hat k_\kappa(\hat q_\xa,\hat q_\xe)  - 2 k_\kappa(\hat q_\xa,\hat q_\xe) \tq_\xa \partial_{\tq_\xa} \tf^\lambda_\kappa(\hat q_\xa,\hat q_\xe)  \Bigr)   \Bigr] , \\
\tI^{\xa\xe}_\kappa &= -\tg_\kappa \frac{v_{d_\xa}v_{d_\xe} }{4d_\xa}   \!\! \int_{0}^{\infty} \! \! \!  d \tq_\xa \, \tq^{d_\xa-1}_\xa \! \! \! \int_{0}^{\infty} \! \!  \! d \tq_\xe \,  \tq^{d_\xe -1}_\xe \,   \frac{\tq^{2}_\xe r(\tq^2)}{\tf^\lambda_\kappa(\hat q_\xa,\hat q_\xe) (\hat l_\kappa(\hat q_\xa,\hat q_\xe))^3} \Bigl[ 
- \Bigl(2 \hat q_\xa^2 -d_\xa (\tq_\xa^2 + \gamma_\kappa \tq_\xe^2) \Bigr) \tf^\lambda_\kappa(\hat q_\xa,\hat q_\xe)  \hat k_\kappa(\hat q_\xa,\hat q_\xe) \nonumber \\ & 
\qquad + (\tq_\xa^2 + \gamma_\kappa \tq_\xe^2) \Bigl(\tf^\lambda_\kappa(\hat q_\xa,\hat q_\xe) \tq_\xa \partial_{\tq_\xa} \hat k_\kappa(\hat q_\xa,\hat q_\xe)  - 2 \hat k_\kappa(\hat q_\xa,\hat q_\xe) \tq_\xa \partial_{\tq_\xa} \tf^\lambda_\kappa(\hat q_\xa,\hat q_\xe)  \Bigr)   \Bigr] , \\
\tI^{\xa \xn}_\kappa &= \tg_\kappa  \frac{v_{d_\xa}v_{d_\xe} }{2d_\xa}   \!\! \int_{0}^{\infty} \! \! \!  d \tq_\xa \, \tq^{d_\xa-1}_\xa \! \! \! \int_{0}^{\infty} \! \!  \! d \tq_\xe \,  \tq^{d_\xe -1}_\xe \,   \frac{\tq^{2} (\partial_{\tq^2} r(\tq^2)) }{\tf^\lambda_\kappa(\hat q_\xa,\hat q_\xe) (\hat l_\kappa(\hat q_\xa,\hat q_\xe))^3} 
\times
\nonumber \\ & \qquad
\Bigl[   \tf^\lambda_\kappa(\hat q_\xa,\hat q_\xe)  \Bigl(2 ( \gamma_\kappa-1) (\hat q_\xa^2 \hat q_\xe^2 / \hat q^2)     \hat l_\kappa(\hat q_\xa,\hat q_\xe)  + \hat q^2 (2 \hat q_\xa^2 -d_\xa (\tq_\xa^2 + \gamma_\kappa \tq_\xe^2) ) \hat k_\kappa(\hat q_\xa,\hat q_\xe) \Bigr)     \nonumber \\ & 
\qquad  + (\tq_\xa^2 + \gamma_\kappa \tq_\xe^2) \Bigl(\tf^\lambda_\kappa(\hat q_\xa,\hat q_\xe)  \Bigl ( \hat q_\xa \partial_{\tq_\xa} \hat l_\kappa(\hat q_\xa,\hat q_\xe)  - \tq^2 \tq_\xa \partial_{\tq_\xa} \hat k_\kappa(\hat q_\xa,\hat q_\xe)  -2 (\hat q_\xa^2 / \hat q^2) \hat l_\kappa(\hat q_\xa,\hat q_\xe) \Bigr )    
\nonumber \\ & \qquad
 + 
(2 \tq^2 \hat k_\kappa(\hat q_\xa,\hat q_\xe) - \hat l_\kappa(\hat q_\xa,\hat q_\xe)) \hat q_\xa \partial_{\tq_\xa} \tf^\lambda_\kappa(\hat q_\xa,\hat q_\xe) \Bigr)
\Bigr] , \\
\tI^{\xe \td}_\kappa &= \tg_\kappa \gamma_\kappa \frac{v_{d_\xa}v_{d_\xe} }{4d_\xe}  \int_{0}^{\infty} \! \! \!  d \tq_\xa \, \tq^{d_\xa-1}_\xa \! \! \! \int_{0}^{\infty} \! \!  \! d \tq_\xe \,  \tq^{d_\xe -1}_\xe \,  \frac{ r(\tq^2)}{ \tf^\lambda_\kappa(\hat q_\xa,\hat q_\xe) (\hat l_\kappa(\hat q_\xa,\hat q_\xe))^3} \Bigl[ 
2 (1- \gamma_\kappa) (\hat q_\xa^2 \hat q_\xe^2 / \hat q^2)   \tf^\lambda_\kappa(\hat q_\xa,\hat q_\xe)  \hat l_\kappa(\hat q_\xa,\hat q_\xe)   \nonumber \\
& \qquad + (\tq_\xa^2 + \gamma_\kappa \tq_\xe^2)  \Bigl(  \tf^\lambda_\kappa(\hat q_\xa,\hat q_\xe)  \hat q_\xe \partial_{\tq_\xe} \hat l_\kappa(\hat q_\xa,\hat q_\xe) - \hat l_\kappa(\hat q_\xa,\hat q_\xe) \hat q_\xe \partial_{\tq_\xe} \tf^\lambda_\kappa(\hat q_\xa,\hat q_\xe) -2 (\hat q_\xe^2 / \hat q^2) \tf^\lambda_\kappa(\hat q_\xa,\hat q_\xe)\hat l_\kappa(\hat q_\xa,\hat q_\xe)  \Bigr)   \Bigr] ,  \\
\tI^{\xe\xa}_\kappa &= -\tg_\kappa \gamma_\kappa \frac{v_{d_\xa}v_{d_\xe} }{4d_\xe}   \!\! \int_{0}^{\infty} \! \! \!  d \tq_\xa \, \tq^{d_\xa-1}_\xa \! \! \! \int_{0}^{\infty} \! \!  \! d \tq_\xe \,  \tq^{d_\xe -1}_\xe \,  \frac{\tq^{2}_\xa r(\tq^2)}{\tf^\lambda_\kappa(\hat q_\xa,\hat q_\xe) (\hat l_\kappa(\hat q_\xa,\hat q_\xe))^3} \Bigl[   - \Bigl(2 \gamma_\kappa \hat q_\xe^2 -d_\xe (\tq_\xa^2 + \gamma_\kappa \tq_\xe^2) \Bigr) \tf^\lambda_\kappa(\hat q_\xa,\hat q_\xe)  \hat k_\kappa(\hat q_\xa,\hat q_\xe)  \nonumber \\ & 
\qquad  + (\tq_\xa^2 + \gamma_\kappa \tq_\xe^2) \Bigl(\tf^\lambda_\kappa(\hat q_\xa,\hat q_\xe) \tq_\xe \partial_{\tq_\xe} \hat k_\kappa(\hat q_\xa,\hat q_\xe)  - 2 k_\kappa(\hat q_\xa,\hat q_\xe) \tq_\xe \partial_{\tq_\xe} \tf^\lambda_\kappa(\hat q_\xa,\hat q_\xe)  \Bigr) \Bigr] , \\
\tI^{\xe\xe}_\kappa &= -\tg_\kappa \gamma_\kappa \frac{v_{d_\xa}v_{d_\xe} }{4d_\xe}   \!\! \int_{0}^{\infty} \! \! \!  d \tq_\xa \, \tq^{d_\xa-1}_\xa \! \! \! \int_{0}^{\infty} \! \!  \! d \tq_\xe \,  \tq^{d_\xe -1}_\xe \,  \frac{\tq^{2}_\xe r(\tq^2)}{\tf^\lambda_\kappa(\hat q_\xa,\hat q_\xe) (\hat l_\kappa(\hat q_\xa,\hat q_\xe))^3} \Bigl[   - \Bigl(2 \gamma_\kappa \hat q_\xe^2 -d_\xe (\tq_\xa^2 + \gamma_\kappa \tq_\xe^2) \Bigr) \tf^\lambda_\kappa(\hat q_\xa,\hat q_\xe)  \hat k_\kappa(\hat q_\xa,\hat q_\xe)  \nonumber \\ & 
\qquad  + (\tq_\xa^2 + \gamma_\kappa \tq_\xe^2) \Bigl(\tf^\lambda_\kappa(\hat q_\xa,\hat q_\xe) \tq_\xe \partial_{\tq_\xe} \hat k_\kappa(\hat q_\xa,\hat q_\xe)  - 2 k_\kappa(\hat q_\xa,\hat q_\xe) \tq_\xe \partial_{\tq_\xe} \tf^\lambda_\kappa(\hat q_\xa,\hat q_\xe)  \Bigr) \Bigr] , \\
\tI^{\xe \xn}_\kappa &= \tg_\kappa \gamma_\kappa \frac{v_{d_\xa}v_{d_\xe} }{2d_\xe}   \!\! \int_{0}^{\infty} \! \! \!  d \tq_\xa \, \tq^{d_\xa-1}_\xa \! \! \! \int_{0}^{\infty} \! \!  \! d \tq_\xe \,  \tq^{d_\xe -1}_\xe \,   \frac{\tq^{2} (\partial_{\tq^2} r(\tq^2)) }{\tf^\lambda_\kappa(\hat q_\xa,\hat q_\xe) (\hat l_\kappa(\hat q_\xa,\hat q_\xe))^3} 
\times
\nonumber \\ & \qquad
\Bigl[   \tf^\lambda_\kappa(\hat q_\xa,\hat q_\xe)  \Bigl(2 (1- \gamma_\kappa) (\hat q_\xa^2 \hat q_\xe^2 / \hat q^2)     \hat l_\kappa(\hat q_\xa,\hat q_\xe)  + \hat q^2 (2 \gamma_\kappa \hat q_\xe^2 -d_\xe (\tq_\xa^2 + \gamma_\kappa \tq_\xe^2) ) \hat k_\kappa(\hat q_\xa,\hat q_\xe) \Bigr)     \nonumber \\ & 
\qquad  + (\tq_\xa^2 + \gamma_\kappa \tq_\xe^2) \Bigl(\tf^\lambda_\kappa(\hat q_\xa,\hat q_\xe)  \Bigl ( \hat q_\xe \partial_{\tq_\xe} \hat l_\kappa(\hat q_\xa,\hat q_\xe)  - \tq^2 \tq_\xe \partial_{\tq_\xe} \hat k_\kappa(\hat q_\xa,\hat q_\xe)  -2 (\hat q_\xe^2 / \hat q^2) \hat l_\kappa(\hat q_\xa,\hat q_\xe) \Bigr )    
\nonumber \\ & \qquad
 + 
(2 \tq^2 \hat k_\kappa(\hat q_\xa,\hat q_\xe) - \hat l_\kappa(\hat q_\xa,\hat q_\xe)) \hat q_\xe \partial_{\tq_\xe} \tf^\lambda_\kappa(\hat q_\xa,\hat q_\xe) \Bigr)
\Bigr] ,
\end{align}
\label{eq:etaint}
\end{subequations}
\end{widetext}
where
\begin{subequations}
\begin{align}
\hat k_\kappa( \hat q_\xa,\hat q_\xe) &= \hat f^\td_\kappa( \hat q_\xa,\hat q_\xe)+ r( \hat q^2) , \\
\hat l_\kappa( \hat q_\xa,\hat q_\xe) &= \hat q^2 ( \hat f^\nu_\kappa(\hat q_\xa,\hat q_\xe)+ r( \hat q^2 )), \\
\hat q^2 &= \hat q_\xa^2 + \hat q_\xe^2 .
\end{align}
\label{eq:etadef}
\end{subequations}
The flow integrals in the LPA' approximation are easily deduced from the above NLO expressions by setting the flowing functions to one according to Eq.\ (\ref{eq:simple}). 
Using the substitution $\hat q_\xa = \hat q \cos\theta $, $\hat q_\xe = \hat q \sin\theta $ 
the double integrals in Eqs.\ (\ref{eq:etaint}) can be further simplified to simple radial integrals 
\begin{subequations}
\begin{align}
\int_{0}^{\infty} \! \! \!  d \tq_\xa \, \tq^{d_\xa-1}_\xa \! \! \! \int_{0}^{\infty} \! \!  \! d \tq_\xe \,  \tq^{d_\xe -1}_\xe \,  \hat{I}(\hat q) &= \frac{v_{d}}{v_{d_\xa}v_{d_\xe} } \int_{0}^{\infty} \! \! d \tq \, \tq^{d-1}  \hat{I}(\hat q),  \\
 \int_{0}^{\infty} \! \! \!  d \tq_\xa \, \tq^{d_\xa+1}_\xa \! \! \! \int_{0}^{\infty} \! \!  \! d \tq_\xe \,  \tq^{d_\xe -1}_\xe \,  \hat{I}(\hat q) &= \frac{v_{d} d_\xa }{v_{d_\xa}v_{d_\xe} d} \int_{0}^{\infty} \! \! d \tq \, \tq^{d+1}  \hat{I}(\hat q),
\end{align}
\label{eq:radialint}
\end{subequations}
where $\hat{I}(\hat q)$ is a radial symmetric integrand.

In the isotropic case $\hat{f}^\xx(\hat q_\xa,\hat q_\xe) \equiv \hat{f}^\xx(\hat q)$ and  $\gamma = 1$. 
 One can then easily check that Eqs.\ (\ref{eq:etaint})  with the help of Eq.\ (\ref{eq:radialint}) simplify to the isotropic expressions (A4) in Refs.\ \cite{Kloss12,Kloss14},
with 
\begin{subequations}
\begin{align}
\tI^{\td \nu}_\kappa &= \tI^{\td \xa}_\kappa +  \tI^{\td \xe}_\kappa , \\
\tI^{\nu \nu}_\kappa &= \tI^{\xa \xa}_\kappa + \tI^{\xa \xe}_\kappa = \tI^{\xe \xa}_\kappa + \tI^{\xe \xe}_\kappa , \\
\tI^{\nu \xn}_\kappa &= \tI^{\xa \xn}_\kappa =  \tI^{\xe \xn}_\kappa , \,\,\, \tI^{\nu \td}_\kappa = \tI^{\xa \td}_\kappa =  \tI^{\xe \td}_\kappa .
\end{align}
\end{subequations}

\renewcommand{\theequation}{D\arabic{equation}}
\section*{Appendix D: On the nature of the anisotropic (A) fixed point}
\setcounter{equation}{0}

\begin{figure}[tb]
  \centering
\includegraphics[width=80mm]{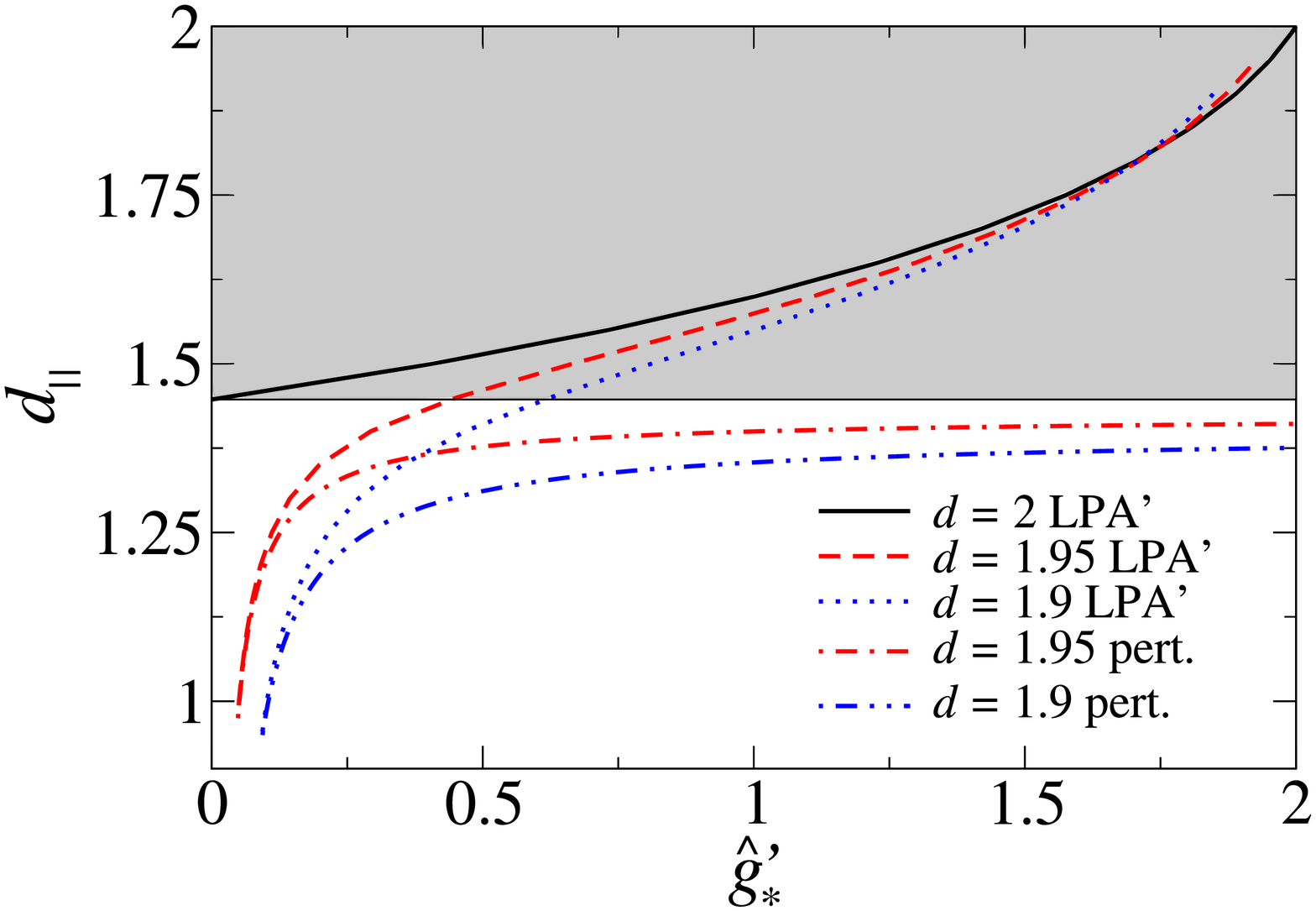}
  \vspace{-4mm}
  \caption{%
(Color online) Coupling constant $\tg'_*$  at the anisotropic A fixed point close to $d = 2$ for different values of the sector dimension $d_\xa$. The LPA' results are obtained from a numerical solution of the NPRG flow equations and  perturbative results (labeled as pert.) from the one-loop expression Eq.\ (\ref{eq:oneloopg}). For $d < 2$ and small $d_\xa$ the A fixed point is weak coupling and is well described by a perturbative treatment.
For larger values of $d_\xa$ the A fixed point becomes strong coupling and the perturbative treatment is no longer justified.
The sector dimension $d_\xa = 1 + 1/\sqrt{5} \approx 1.45$ stands as the boundary beyond which  the one-loop treatment for A breaks down, separating the weak coupling (white) from the strong coupling regime (shaded in grey).
}
  \label{fig:Apert}
\end{figure}

We start with the $\beta$ functions Eq.\ (\ref{eq:betaflow}) in the perturbative limit Eq.\ (\ref{eq:eta1}). As mentioned before, in the limit of small coupling $\tg$ the $\beta$ functions coincide with the perturbative ones studied by TF \cite{taeuber02}.
Let us define $\epsilon = 2 - d$ with $\epsilon > 0$. For small $\epsilon$ and
$\hat{g} \sim \mathcal{O}(\epsilon)$ the one-loop $\beta$ functions become
\begin{subequations}
\begin{align}
\beta_{\hat{g}} &=  - \epsilon \tg + \beta_0 \tg^2  + \mathcal{O} (\epsilon^3) ,
\label{eq:betapertg} \\
\beta_{\gamma} &=  \tg \gamma (\gamma\! -\! 1) (2 - \Delta + \gamma (2 + \Delta)) A(d)\!  +\! \mathcal{O} (\epsilon^2) ,
\end{align}
\label{eq:betapert}
\end{subequations}
where
\begin{equation}
\beta_0 =\! (8 \! -\! 16 \Delta\! -\! 2 \Delta^2 \! - \!4 \gamma (8\! - \!2 \Delta\! -\! \Delta^2) \!+ \!\gamma^2 (2 \!- \!\Delta)^2) A(d) . \nonumber
\end{equation}
The  $\beta_{\gamma}$ function has three zeros (see  Sec.\ \ref{sec:results}):  $\gamma_* = 1$ (isotropic case), $\gamma_* = 0$ (unitary case) and $\gamma_* = (\Delta - 2) / (\Delta + 2)$ (anisotropic case). Substituting the  anisotropic fixed point value for $\gamma$ into the first $\beta$ function Eq.\ (\ref{eq:betapertg}) we obtain
\begin{equation}
\beta_A (\tg) = - \epsilon \tg + \frac{8 A(d) (4 - 5 (d - 2 d_\xa)^2)}{(d - 2 (1 + d_\xa))^2} \tg^2 + \mathcal{O} (\epsilon^3) .
\end{equation}
This function has the nontrivial fixed point
\begin{equation}
\tg'_* = - \frac{v_d }{4 A(d)} \frac{(d-2) (d - 2 (1 + d_\xa))^2 }{8 (4 - 5 (d - 2 d_\xa)^2)} .
\label{eq:oneloopg}
\end{equation}
In order to stay in the perturbative regime where $\hat{g}_* \sim \mathcal{O}(\epsilon)$ is valid, the dimension $d_\xa$ has to verify $d_\xa < 1 + 1/\sqrt{5}$.
For $d < 2$ and depending on the splitting in the sector dimensions, the A fixed point therefore changes  from weak coupling (for $d_\xa < 1 + 1/\sqrt{5}$) to become strong coupling (for $d_\xa \geq 1 + 1/\sqrt{5}$), see Fig.\ \ref{fig:Apert}.

Let us mention that for $d_\xa \geq 1 + 1/\sqrt{5}$ and $d \geq 2$ the perturbative treatment can be used in order to analyze the A fixed point as long as at least two loop contributions are included and $d_\xa - 1 - 1/\sqrt{5} \sim \sqrt{\epsilon}$.


Finally, a similar analysis is possible for the uniaxial U fixed point. The critical sector dimension is $d_\xa = \sqrt{5}-1$ in that case.

\renewcommand{\theequation}{E\arabic{equation}}
\section*{Appendix E: Gauge Symmetry of the uniaxial AKPZ equation}
\setcounter{equation}{0}

The AKPZ action in the uniaxial case ($\gamma = 0$) is invariant under the generalized gauged shift
\begin{equation}
 h'(t,\vx)=h(t,\vx)+f(t,\vec{x}_\xe) ,
\end{equation}
except for the two terms 
\begin{equation}
\delta S = \int_{\bf{x}} \left\{ \tilde{h}({\bf{x}}) ( \partial_t f(t,\vec{x}_{\xe}) - \nu_{\xe} \nabla^2_{\xe} f(t,\vec{x}_\xe) )  \right\} ,
\end{equation}
which variation is linear in the field.
Following the same line of reasoning as  Ref.\ \cite{Canet11a}, one deduces that 
 \begin{equation}
 \int_{\vec{x}_\xa} \frac{\delta \Gamma}{\delta h} = \int_{\vec{x}_\xa} \frac{\delta S}{\delta h}  = 0 ,
\end{equation}
or equivalently
\begin{equation}
 \Gamma^{(n,m)} (\vec{p}_\xa = 0,\ldots) =( i \omega + \nu_\xe p^2_\xe )\delta_{m1} \delta_{n1} .
\end{equation}
This implies that $\eta_\kappa^\xe = 0$ since $\nu_\kappa^\xe$ is not renormalized  in the flow.

\end{appendix}

\bibliography{biblioKPZ}
\bibliographystyle{apsrev4-1}

\end{document}